\documentclass[12pt]{article}

\pdfoutput=1
\usepackage[pdftex]{color}
\usepackage{amssymb}
\usepackage{amsthm}
\usepackage{amsmath}
\usepackage{latexsym}
\usepackage{epsfig}
\usepackage{amscd}
\usepackage{graphicx}
\usepackage[pdftex, colorlinks=true, citecolor=green]{hyperref}
\usepackage{lscape}
\usepackage{setspace}
\usepackage{cancel}
\usepackage{enumerate}
\usepackage{mathrsfs}
\usepackage{cite}

\renewcommand{\proof}[1]{\noindent \normalfont{\textbf{Proof.} \  \  #1 \hfill $\Box$}\\}

\newtheorem{thm}{Theorem}[section]
\newtheorem{prop}[thm]{Proposition}
\newtheorem{lemma}[thm]{Lemma}
\newtheorem{corol}[thm]{Corollary}
\newtheorem{defn}[thm]{Definition}
\newtheorem{conj}[thm]{Conjecture}

\newcommand{\ds}{\displaystyle}

\begin{document}

\date{}

\title{Neural network spectral robustness under perturbations of the underlying graph}

\author{Anca R\v{a}dulescu$^1$}

\maketitle

$^1$Department of Mathematics, State University of New York at New Paltz, New Paltz, NY 12561 \\

\hspace{2mm} phone: (845)257-3532; fax: (845)257-3571; email: \emph{radulesa@newpaltz.edu}\\

\maketitle

\begin{abstract}
\noindent Recent studies have been using graph theoretical approaches to model complex networks (such as social, infrastructural or biological networks), and how their hardwired circuitry relates to their dynamic evolution in time. Understanding how configuration reflects on the coupled behavior in a system of dynamic nodes can be of great importance, for example in the context of how the brain connectome is affecting brain function.  However, the connectivity patterns that appear in brain networks, and their individual effects on network dynamics, are far from being fully understood.\\

\noindent We study the connections between edge configuration and dynamics in a simple oriented network composed of two interconnected cliques (representative of brain feedback regulatory circuitry). In this paper, our main goal is to study the spectra of the graph adjacency and Laplacian matrices, with a focus on three aspects in particular: (1) the sensitivity/robustness the spectrum in response to varying the intra and inter-modular edge density, (2) the effects on the spectrum of perturbing the edge configuration, while keeping the densities fixed and (3) the effects of increasing the network size. We study some tractable aspects analytically, then simulate more general results numerically. This paper aims to clarify, from analytical and modeling perspectives, the underpinnings of our related work, which further addresses how graph properties affect the network's temporal dynamics and phase transitions.\\

\noindent We propose that this type of results may be helpful when studying small networks such as macroscopic brain circuits. We suggest potential applications to understanding synaptic restructuring in learning networks, and the effects of network configuration to function of emotion-regulatory neural circuits.
\end{abstract}

\noindent {\bf Keywords:} oriented graph; edge density; adjacency matrix; graph Laplacian; eigenvalue spectrum; robustness; neural network; brain connectome.

\clearpage
\section{Introduction}

\subsection{Network architecture and brain connectivity}

The study of networks has been the subject of great interest in recent research. Many natural systems are organized as networks, in which the nodes (be they cells, individuals or web servers) interact in a time-dependent fashion.

One of the particular points of interest has been the question of how the hardwired \emph{structure} of a network (its underlying graph) affects its \emph{function}, for example in the context of optimal information storage or transmission between nodes along time~\cite{bullmore2009complex}. It has been hypothesized that there are two key conditions for optimal function in such networks: a well-balanced adjacency matrix (the underlying graph should appropriately combine robust features and random edges) and well-balanced connection strengths, driving optimal dynamics in the system.  A subsequent line of study is to understand the effects of connectivity patterns on the temporal behavior of the network -- seen as a dynamical system, in which the node-variables are coupled according to a connectivity scheme that obeys certain deterministic constrains, but also incorporates random aspects. One can then investigate how the phase space dynamics (and the phase transitions that the system undergoes under perturbation) are affected when perturbing the underlying adjacency graph.

Recent studies have used graph theoretical approaches to investigate brain networks, not only in the context of learning, memory formation and cognitive performance, but also to understand more general organizational and functional principles used by the brain~\cite{bullmore2009complex,sporns2011non,sporns2002graph}. With nodes and edges defined at various scales, according to different empirical modalities~\cite{sporns2010networks}, these studies support certain generic topological properties of brain architecture, such as modularity, small-worldness, the existence of hubs and other connectivity density patterns~\cite{he2010graph}. These properties, if proven consistent with physiological, behavioral or genetic factors, may provide us with a better understanding of neural processes, and may be effective as biomarkers for behavioral traits or neuropsychiatric conditions.

In the context of the brain, a network of nodes connected by oriented weighted edges may constitute a valid representation of neural architecture at more than one spatio-temporal scale ( the brain ``fractal'' possibly reusing similar organizational and optimization principles at multiple complexity levels). For example, one may think of the nodes as individual neurons connected by synapses, whose activity (e.g., measured as variations in membrane potential) is determined by the external input together with the pattern and strengths of synaptic coupling. Such a microscopic framework would be best placed in connection with \emph{in vitro} recordings from single cells. At a coarser level, one may view a node as a functional population of neurons, whose activity (e.g., measured as mean field firing rate) is determined by the external input, as well as its mean field connections from other -- excitatory or inhibitory -- populations. At the top macroscopic level, consistent with imaging techniques such as MRI, or EEG, one may think of the nodes as anatomical brain regions (e.g., amygdala, prefrontal cortex), whose activity (e.g., measured as an ERP, or BOLD signal) is determined by the external stimulus and the inter-regional connectivity patterns and connection strengths. The way in which various parts of the brain (from the micro-scale of neurons to the macro-scale of functional regions) are wired together is one of the great scientific challenges of the 21st century, currently being addressed by large-scale research collaborations, such as the Human Connectome Project~\cite{toga2012mapping,craddock2013imaging}. While the general aim of our own research is to study the relationship between a network's hardwired circuitry and its dynamics, this paper focuses primarily on understanding some of the features of the underlying graph, and to a lesser extent on investigating their potential to further affect the dynamic vulnerability or robustness or the system (which is the subject of a related paper~\cite{Chaos2015}).

One thought of potential importance to us is that, while the brain itself is a gigantic and relatively densely-connected network of billions of neuron-nodes (each receiving and providing input to tens of thousands of other nodes), it may be both realistic and computationally advantageous to view the brain as a highly hierarchic network, in which the behavior of each one ``node'' at a certain complexity level integrates the behavior of a collection of lower-level nodes.
Hence, at each complexity level, the size of the networks we need to study experimentally, represent theoretically or simulate numerically may be in fact relatively small (a few hundred nodes). For example, at the macroscopic level compatible with imaging techniques in humans, a small region such as the amygdala is (within typical fMRI acquisitions parameters) as large as 100-200 voxel-nodes. For relatively small networks, the traditional large size limit results obtained in random graph theory may no longer apply directly, and new approaches need to be created to extend the results (see Section~\ref{general} for a more detailed discussion).

\subsection{Brain function and graph theory of the connectome}
\label{connectome}

Recent studies have attempted to identify dynamic patterns (such as return to baseline after perturbation, signal complexity, proximity of the system to a critical dynamic range) from imaging time series in humans, and interpret them in connection with connectivity patterns between the brain's  coupled components. In large-scale model networks constructed from neuroimaging data on the brain, modeling can address different types of connectivity: structural connectivity, for anatomical links; functional connectivity, for undirected statistical dependencies; and effective connectivity, for directed causal relationships among distributed responses~\cite{friston2013structural}.

A lot of effort has been invested recently towards developing and using graph-theoretical network measures in conjunction with statistical methods, in order to identify the effects of abnormal connectivity patterns on the efficiency of brain function.  By applying graph theoretical measures of segregation (e.g., clustering coefficient, motifs, modularity, rich clubs), integration (e.g., distance, path length, efficiency) and influence (e.g., node degree, centrality), various studies have been investigating the sensitivity of systems to removing/adding nodes or edges to different places in the network structure.

Working with empirical data, such measures have been used to understand behavioral impairments in subjects with compromised connectivity due to existing lesions~\cite{corbetta2012functional}, or group differences between healthy controls and patients with mental illnesses associated with abnormal feedback circuitry. When looking for differences between the connectomes of healthy human controls and those of patients with various neuropsychiatric illnesses, empirical investigators have used tractography, or resting state functional MRI data, or both. For example, a graph theory based network approach to the tractography-derived brain connectome detected successfully network integration deficits in 25 euthymic bipolar patients when compared to 24 age and gender matched healthy controls~\cite{leow2013impaired,gadelkarim2013investigating}, as well as in 42 subjects with DSM-IV major depression versus 47 matched healthy controls~\cite{gadelkarim2012framework}.

Network analyses were also used efficiently in conjunction with functional connectivity measured via resting state fMRI (rsfMRI). A recent study~\cite{fekete2013combining} used a publicly available rsfMRI subject sample including 18 males, out of which 8 schizophrenia patients, to compute functional connectivity metrics. The data sets were mined for global (e.g., characteristic path length, clustering coefficient, small-world ratio parameter) and local  (e.g., nodal-betweenness centrality, nodal path length, nodal clustering coefficient) network measures. These measures were then used as features for several support vector based classifiers, providing means to select the network model that would optimally differentiate between subjects (e.g., lower small world ration in schizophrenia patients), with a success rate of 94-100$\%$.

Clearly, the development of multiple imaging modalities has made it increasingly feasible to simultaneously capture hardwired and temporal aspects of the connectome, and to understand them together using the data in conjunction with graph theoretical methods. New computational techniques  integrate multimodal data from resting state fMRI and from the whole brain tractography-derived connectome, increasing the power to detect group differences in brain connectivity~\cite{ajilore2013constructing}.

However, no matter how well-designed or statistically powerful, purely empirically-based analyses cannot explain in and off themselves the mechanisms by which connectivity patterns actually act to change the system's dynamics, and thus the observed behavior. Substantial research effort is being directed towards constructing an underlying network model that is tractable theoretically or numerically, and which could therefore be used \emph{ in conjunction} with the data towards interpreting the empirical results, and for making further predictions. To this aim, the theoretical dependence of dynamics on connectivity (e.g., in the context of stability and synchronization in networks of coupled neural populations) has been investigated both analytically and numerically, in a variety of contexts -- from biophysical models~\cite{gray2009stability} to simplified systems~\cite{siri2007effects}. These analyses revealed a rich range of potential dynamic regimes and transitions~\cite{brunel2000dynamics}, shown to depend as much on the coupling parameters of the network as on the arrangement of the excitatory and inhibitory connections~\cite{gray2009stability}. The construction of a realistic, data-compatible computational model has been subsequently found to present many difficulties related to dimensionality (in both phase and parameter space), to seamlessly patching the multiple network scales, to appropriate inclusion of  stochastic/noisy aspects.

In our own previous work~\cite{radulescu2013network}, we focused on addressing some of these problems; we chose to use a simple graph-theoretical model as a formal framework to study how network density can affect the complexity of signal outputs. Simple and quite  general, this setup informed successfully our human imaging results in the circuit regulating human emotion (see Section~\ref{application} for a more detailed description of the modeling results). The clinical promise of this model motivated our effort to gain a better understanding of the theoretical properties behind some of the more important (and sometimes counterintuitive) results suggested by our computational model. Among these are the robustness of the coupled dynamics to certain changes in the network architecture and its vulnerability to others, as well as the differences between updating connection strengths versus perturbing connection density or geometry. Because of its simple and general set-up, and of its demonstrated applicability, we found it informative to start precisely with the network described in this previous work, which, in this simple form, opens questions on (1) properties of a bimodular graph with variable edge distribution (e.g, of the adjacency and Laplacian spectra, which we study in this paper), and (2) the dependence of dynamics on coupling parameters in a network with variable architecture (addressed in a separate paper~\cite{Chaos2015}). In the ongoing iterations of this work are also studying other architectures and extensions.

\subsection{Network dynamics from spectral measures}

\noindent {\bf The adjacency spectrum.} A variety of studies have examined random graphs with a general given expected degree distribution, and have established bounds or other descriptions of their adjacency spectra. While it is well know that the largest eigenvalue of a graph's adjacency matrix is determined by its maximum degree $m$ together with the weighted average $\tilde{d}$ of the squares of the expected degrees~\cite{chung2006complex}, recent work on random matrices has delivered more accurate estimates. For example, Chung et al.~\cite{chung2003spectra} have investigated an ensemble of random uncorrelated, non-oriented networks, and found that, in the large $N$ limit, the expected largest eigenvalue is determined by the ratio of the second to first moment of the average degree distribution $\langle d^2 \rangle / \langle d \rangle$, together with the expected largest degree $d_\text{max}$. More generally, for directed (oriented) networks without edge degree correlations, a first order approximation to the leading eigenvalue is given by $\langle d^\text{in} d^\text{out} \rangle / \langle d \rangle$, where $d^\text{in}$ and $d^\text{out}$ are respectively the in and out-degrees of the graph, and $\langle d^\text{in} \rangle = \langle d^\text{out} \rangle = \langle d \rangle$~\cite{restrepo2007approximating}.

It is therefore clear that the in/out degrees, as well as their correlations, have crucial effects on the leading eigenvalue. In general, a graph's defining feature is its distribution of edges. Among other properties, edge density, edge clustering and presence of hubs, have been intensely studied. Detecting and interpreting the \emph{ modularity} of a network (i.e., the presence of community structures within the graph, defined as densely connected groups of nodes, with sparser inter-group connections) has been recently of particular interest~\cite{chauhan2009spectral,nadakuditi2012graph,nadakuditi2013spectra,sarkar2013spectral}. Whether the graph represents the architecture of a social~\cite{gilbert2011communities}, climate~\cite{donges2009complex}, transportation~\cite{zanin2013modelling} or disease~\cite{barabasi2011network,van2011n,supekar2008network} network, modularity reflects into adjacency properties of the network, controlling the structural and functional properties, and implicitly the temporal behavior of the system.\\

\noindent {\bf The graph Laplacian spectrum.} The Laplacian matrix $L$ of a graph is defined as the difference between the node degree matrix and the adjacency matrix.  In the case of directed graphs, either the indegree or outdegree can be used, depending on the application. Laplacian dynamics is perhaps the most studied representation of networked systems, and is also known as the consensus protocol~\cite{olfati2007consensus}, in which, the network aims to reach agreement on a certain quantity of interest. Although this model has been explored in more elaborate contexts~\cite{olfati2004consensus,rahmani2009controllability}, in its simplest form the dynamics of each node is driven by the sum of differences between its own state and its neighbors' states, as defined by the adjacency graph. Then, the dynamic evolution of the entire system can be appropriately captured by the linear equation: $\dot{x}(t) = -Lx(t)$. \\

\noindent While the consensus protocol has attracted a lot of attention and effort~\cite{wu2013control}, it is not a complete representation of all the recent work on networked dynamic systems. For example, relative sensing networks are an important class of systems whose control has been described both using their incidence matrix~\cite{smith2007closed}, as well as, more completely, in terms of spanning trees in the connection topology~\cite{sandhu2005relative}. In fact, the dynamical stability of certain networks seems to remain most successfully defined in terms of quantities derived from the eigenspectrum of the adjacency matrix~\cite{small2012stability}.

In  our own work, we considered a bimodal oriented network of coupled nodes, each acting as Wilson-Cowan type nonlinear oscillator~\cite{Chaos2015}. Even for such a network, one cannot expect either adjacency or the Laplacian spectrum to be fully predictive of the system's dynamics. Indeed, both cospectral graphs and Laplacian cospectral graphs may produce different phase and parameters-space behavior in the corresponding system (examples of this correspondence are shown in Appendix A). A stronger requirement for the graphs to be \emph{ isomorphic} would most likely lead to identical coupled dynamics; but, while isomorphic graphs are cospectral and Laplacian cospectral respectively, the converse is not true in either case~\cite{barghi2009non,zelazo2008analysis}. These being said, however, both adjacency and graph Laplacian matrices have properties which reflect into the network dynamic behavior. \\

\noindent In this study, we focus on understanding the spectral properties of a bimodular oriented network. More precisely, we consider oriented graphs with two interconnected modules $X$ and $Y$, each composed of $N$ nodes. Within both $X$ and $Y$, the edge density is fixed to the same fraction $\gamma$ (out of the possible maximum of $N^2$). The density of the $X$-to-$Y$ edges is fixed to a fraction $\alpha$ of the $N^2$ possible $X$-to-$Y$ connections, and the density of the $Y$-to-$X$ edges is fixed to $\beta$ of the $N^2$ possible $Y$-to-$X$ connections. The parameters $\alpha, \beta$ and $\gamma$ can take any values of the form $\frac{k}{N^2} \in [0,1]$, where $k$ is an integer between zero and $N^2$ (not necessarily requiring that $\gamma > \alpha, \beta$). In this setup, when $\gamma=0$ the modules are totally disconnected, and when $\gamma=1$ the modules are fully connected (\emph{ cliques}). Most of this paper is dedicated to studying interconnected cliques.

As discussed in Section~\ref{connectome}, this graph structure was used in previous work as a schematic architectural representation of a neural circuit, in which $X$ and $Y$ represent the excitatory, respectively inhibitory modules of a neural feedback loop, so that $X$ projects to $Y$ through a fraction $\alpha$ of excitatory connections, and $Y$, in turn, modulates $X$ through a fraction $\beta$ of feedback, inhibitory connections. In such a circuit, the overall connectivity density may remain constant during a cognitive process such as learning, even though the network may exhibit high plasticity, and constantly inspect a variety of edge geometry combinations. Throughout the process, the connectivity profile is constantly remodeled, with existing connections being silenced or disappearing, while other, new connections being created or activated.

The adjacency matrix of such an oriented graph is a $2N \times 2N$ binary block matrix of the form: $\ds {\bf T}= \left[ \begin{array}{c|c} {\bf P} & {\bf A} \\ \cline{1-2} {\bf B} & {\bf Q}  \end{array} \right]$, where the blocks ${\bf P}$ and ${\bf Q}$ have a fixed fraction $\gamma$ of $1$ versus $0$ entries (i.e., edge density), while ${\bf A}$ and ${\bf B}$ have densities $\alpha$ and $\beta$, respectively. Here, we study the sensitivity and robustness properties of the adjacency and Laplacian spectra for our specific class of oriented graphs. We focus in particular on understanding, for increasing size $N$, how the eigenvalues are perturbed (1) when changing the density profile $(\alpha,\beta, \gamma)$ and (2) when changing only the edge distribution, while keeping densities fixed. We use a combination of analytical and numerical methods to understand the distribution (mean and standard deviation) of each eigenvalue in the adjacency and Laplacian eigenspectrum. In a separate paper (briefly previewed in Section~\ref{application}), we investigate the connections  between graph properties and the dynamics of a corresponding system of coupled node-oscillators.\\

\noindent Our work henceforth is organized as follows. In the following two sections, we study properties of both the adjacency and the Laplacian matrix of the graph. In Section~\ref{adjacency}, we focus on the behavior and robustness of the adjacency spectrum when changing the edge density and configuration. (In the text, we restrict ourselves to the case of two interconnected cliques, $\gamma=1$. However, in Appendix B we relax the full connectedness requirement to $\gamma \leq 1$, and we analyze how the properties of the spectrum change with the trimming of intra-modular edges.) In Section~\ref{Laplacian}, we investigate numerically, by looking at increasing network sizes and variable edge densities, whether the same robustness is characteristic to the spectrum of the graph Laplacian. In Section~\ref{discussion}, we put our results in the context of the existing work on eigenspectra of random graphs. As a preview to our subsequent work in~\cite{Chaos2015}, we briefly explore connections with the temporal behavior of a coupled dynamical system, and discuss the feasibility of dynamic classification based on classes of adjacency or Laplacian spectra. Finally, we present an existing application of using adjacency patterns to quantify efficiency of feedback in a brain circuit, and further discuss the significance of our results in light of neural connectivity and learning plasticity.


\section{Dependence of adjacency spectrum of edge density and network size}
\label{adjacency}

\begin{figure}[h!]
\begin{center}
\includegraphics[width=.6\textwidth]{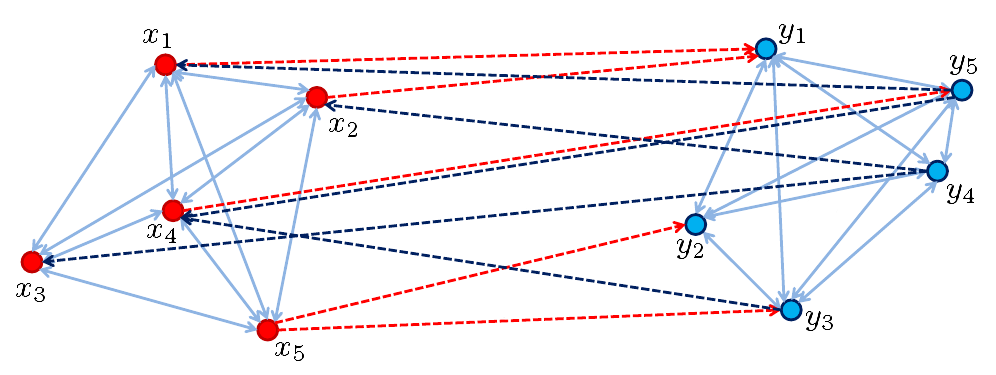}
\end{center}
\caption{ \emph{ \footnotesize {\bf Schematic representation of the network for $N=5$ nodes per module, as used in our application.} Module $X$ is shown on the left; module $Y$ is shown on the right; they are both fully-connected, local sub-graphs of the full network. The dotted red arrows represent the $X$-to-$Y$ connections, and the dotted blue arrows represent the $Y$-to-$X$ connections, generated randomly for low connectivity densities $\alpha=\beta=5/25=0.2$, to maintain clarity of the illustration.}}
\vspace{2mm}
~\cite{Chaos2015}
\label{network}
\end{figure}

We consider the particular case of fully-connected modules $X$ and $Y$ (cliques, see Figure \ref{network}). Then the diagonal blocks of the adjacency matrix ${\bf T}$ are ${\bf P}={\bf Q}={\bf M}$ (where ${\bf M}$ is the appropriate size matrix with all entries equal to one). Note that this scenario includes self-loops at all nodes; eliminating loops is equivalent to subtracting the identity from the adjacency matrix, with the only effect of shifting all the eigenvalues, and preserving the eigenvectors. The off-diagonal $N \times N$ blocks ${\bf A}$ and ${\bf B}$ are binary matrices, with fractions $\alpha$ and respectively $\beta$ of ones.

By discussing the effects of \emph{ edge density} we mean analyzing how the spectrum of ${\bf T}$ changes when the values of $\alpha$ and $\beta$ are varied; we will represent these changes in the form of surface plots with respect to pairs $(\alpha, \beta) \in [0,1]^2$. By discussing the effects of \emph{ geometry} we mean understanding the effects on the spectrum of the edge configuration, under the constraint of fixed densities $\alpha$ and $\beta$. We measure these effects by estimating the mean and standard deviation of the eigenvalues of ${\bf T}$ over the edge geometries admissible by any fixed density pair $(\alpha, \beta)$.

\noindent Let's call $\lambda_j$, $j=\overline{1,2N}$ the eigenvalues of ${\bf T}$, ordered in decreasing order of their magnitudes: $\lvert \lambda_1 \rvert \geq \lvert \lambda_2 \rvert \geq \hdots \geq \lvert \lambda_{2N} \rvert$. Let us notice here that, while $\lambda_1$ is guaranteed to be real by the Perron-Frobenius Theorem, the other eigenvalues are in general complex. We will be referring to $\lambda_1$ and $\lambda_2$ as the two leading eigenvalues of ${\bf T}$.

\begin{defn} For fixed $0 \leq \alpha,\beta \leq 1$, we call ${\cal D}^{\alpha,\beta}$ the distribution of $2N \times 2N$ adjacency matrices ${\bf T}$ with off-diagonal blocks ${\bf A}$ and ${\bf B}$ having densities $\alpha$ and $\beta$, respectively. We call ${\cal L}_j^{\alpha,\beta}$ the corresponding distribution of each of the eigenvalue real parts $Re(\lambda_j)$ (with $j=\overline{1,2N}$).
\end{defn}

\noindent It is easy to see that the cardinality $\ds \lvert {\cal D}^{\alpha,\beta} \rvert = C_{N^2}^{\alpha N^2} C_{N^2}^{\beta N^2}$. While in general the exact eigenvalues of ${\bf T}$ depend on the representative ${\bf T} \in {\cal D}^{\alpha,\beta}$ (i.e., on the actual exact positions of the $1$'s within the blocks ${\bf A}$ and ${\bf B}$), all ${\cal L}_j^{\alpha,\beta}$ are trivial on the boundary (i.e., for $\alpha$ or $\beta$ in $\{0,1\}$).

\begin{lemma}
\label{bound1}
Fixing $\alpha=1$ fixes the eigenvalues of ${\bf T}$, so that $\lvert {\cal L}_j^{1,\beta} \rvert = 1$, for all $j=\overline{1,2N}$. More precisely, the eigenvalues of any ${\bf T} \in {\cal D}^{1,\beta}$ are given by (from largest to smallest in absolute value): $\ds \lambda_1=N+N\sqrt{\beta}$, $\ds \lambda_2=N-N\sqrt{\beta}$, and $\lambda_3 = \hdots = \lambda_{2N} = 0$. Similarly, for $\beta=1$, the eigenvalues of any ${\bf T} \in {\cal D}^{\alpha,1}$ are given by $\ds \lambda_1=N+N\sqrt{\alpha}$, $\ds \lambda_2=N-N\sqrt{\alpha}$, and $\lambda_3 = \hdots = \lambda_{2N} = 0$. \\
\end{lemma}

\proof{We calculate directly, for ${\bf T} \in {\cal D}^{1,\beta}$, the eigenvalues $\lambda$ and eigenvectors $\left[ \begin{array}{c} {\bf V} \\ {\bf W}  \end{array} \right]$ (where ${\bf V} = [v_1,\hdots,v_n]^t$ and ${\bf W}=[w_1,\hdots,w_n]^t$).

$$\left[ \begin{array}{c|c} {\bf M} & {\bf M} \\ \cline{1-2} {\bf B} & {\bf M}  \end{array} \right] \left[ \begin{array}{c} {\bf V} \\ {\bf W}  \end{array} \right] = \lambda \left[ \begin{array}{c} {\bf V} \\ {\bf W}  \end{array} \right]$$

\noindent Call $\ds \Sigma_v = \sum_{j=1}^{N}{v_j}$ and $\ds \Sigma_w = \sum_{j=1}^{N}{w_j}$, and ${\bf B}_j$= the $j$-th row of the block matrix ${\bf B}$, with $\varphi_j({\bf B})$ being the number of $1$s in that row. We then have that:\\

$\ds \Sigma_v + \Sigma_w = \lambda v_j$, for all $j=\overline{1,N}$ and\\

$\ds {\bf B}_j{\bf V} + \Sigma_w = \lambda w_j$, for all $j=\overline{1,N}$.\\

\noindent If $\lambda \neq 0$, then $v_1=v_2=\hdots=v_n=v$, implying that $\Sigma_w = (\lambda-N)v$. It follows that: $\ds \varphi_j({\bf B}) v + \Sigma_w = \lambda w_j$ for all $j=\overline{1,N}$. By summing up, and using the fact that $\sum_{j=1}^{N}{\varphi_j({\bf B})} = N^2 \beta$, we get:\\

$\ds N^2 \beta v + N(\lambda-N)v = \lambda (\lambda-N)v$\\

\noindent Clearly $v \neq 0$, otherwise $w_j=0$ for all $j$, and $\left[ \begin{array}{c} {\bf V} \\ {\bf W}  \end{array} \right] = \left[ \begin{array}{c} {\bf 0} \\ {\bf 0}  \end{array} \right]$. We then have that: $(\lambda-N)^2=N^2 \beta$, hence $\lambda = N \pm N \sqrt{\beta}$.\\

\noindent In conclusion: any matrix ${\bf T} \in {\cal D}^{1,\beta}$ has one largest eigenvalue $\lambda_1 = N+N\sqrt{\beta}$, with eigenvector given by $v_j=v=N+N\sqrt{\beta}$, $w_j=\varphi_j({\bf B})+\sqrt{\beta}$, and a second largest eigenvalue $\lambda_2 = N-N\sqrt{\beta}$, with eigenvector given by $v_j=v=N-N\sqrt{\beta}$, $w_j=\varphi_j({\bf B})-\sqrt{\beta}$. The rest of $2N-2$ eigenvalues are zero. Note that, in the case of $\beta=1$, then $\lambda_1=2N$ and $\lambda_2=0$ as well.
}

\begin{lemma}
\label{bound0}
Fixing $\alpha=0$ fixes the eigenvalues of ${\bf T}$, so that $\lvert {\cal L}_j^{0,\beta} \rvert = 1$, for all $j$. The eigenvalues of any ${\bf T} \in {\cal D}^{0,\beta}$ are given by: $\ds \lambda_1=N$, $\lambda_2 = \hdots = \lambda_{2N} = 0$. Similarly, for $\beta=0$, the eigenvalues of any ${\bf T} \in {\cal D}^{\alpha,0}$ are given by $\ds \lambda_1=N$, $\lambda_2 = \hdots = \lambda_{2N} = 0$.\\
\end{lemma}

\proof{Similar to that of Lemma~\ref{bound1}.}

\noindent Clearly, the distributions ${\cal L}_j^{\alpha,\beta}$ are not trivial in general. If we restricted our interest to finding only the leading eigenvalue of the matrix $\ds {\bf T}= \left[ \begin{array}{c|c} {\bf M} & {\bf A} \\ \cline{1-2} {\bf B} & {\bf M}  \end{array} \right]$, there are a variety of existing tools to assist us. However, even the computations involved in a task such as expanding the powers ${\bf T}^k$ (equivalent to finding all paths of length exactly $k$ in the graph), or in approximating the leading eigenvalue using perturbation theory, become very complex quite fast (see Section~\ref{perturbation} and Appendix B). It is in this light that, at this point, we first proceed numerically to support a few conjectures.

\begin{figure}[h!]
\begin{center}
\includegraphics[width=.9\textwidth]{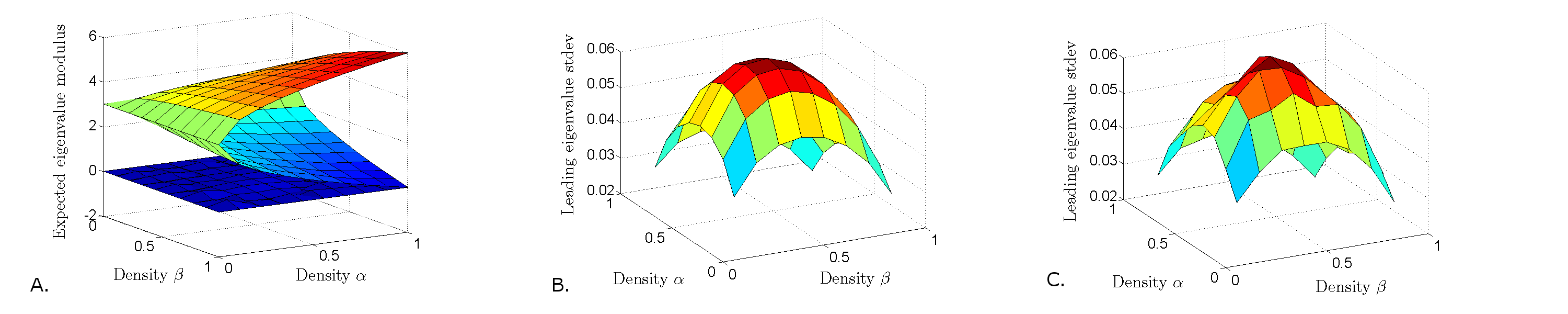}
\end{center}
\caption{ \emph{ \footnotesize {\bf Mean and standard deviation for the leading eigenvalue of ${\bf T}$, for $N=3$}, as functions of the densities $\alpha$ and $\beta$. {\bf A.} For each pair $(\alpha,\beta)$, the mean of the leading eigenvalue real part was calculated over all ${\bf T} \in {\cal D}^{\alpha,\beta}$ (i.e., over all possible combinatorial configurations with the given densities). {\bf B.} For each pair $(\alpha,\beta)$, the corresponding standard deviation was calculated over all combinatorial configurations in each ${\cal D}^{\alpha,\beta}$. {\bf C.} For each pair $(\alpha,\beta)$, the standard deviation of the leading eigenvalue was also calculated using a sample of the distribution,  obtained by choosing randomly $2500$ configurations for ${\bf T}$.}}
\label{eigenvalues3}
\end{figure}

Our goal is to obtain descriptions of ${\cal L}_j^{\alpha,\beta}$ for all values of $\alpha,\beta \in (0,1)$; in particular, we want to estimate their means and standard deviations, and observe how these depend on the values of $\alpha$ and $\beta$ and on the size $N$ of the network. For small network sizes ($N \leq 4$), the mean and standard deviation of the entire distribution ${\cal L}_j^{\alpha,\beta}$, for each $\alpha$, $\beta$ and $j$, can be computed directly quite efficiently (see Figure \ref{eigenvalues3}a and b). However, for larger values of $N$, the factorial increase in the distribution size makes inspecting all configurations computationally very expensive (e.g., for $N=5$ and $\alpha=\beta=12/25$, we have $\lvert {\cal D}^{\alpha,\beta} \rvert = (C_{25}^{12})^2 \sim 10^{13}$ configurations, although some will produce identical spectra). So, for larger $N$s, we estimated the means and standard deviations based on a sample $\mathscr{S} \subset {\cal D}^{\alpha,\beta}$ of the distribution. Figures~\ref{eigenvalues3}b and c show a comparison between the whole-distribution and sample-based computations of the standard deviation for ${\cal L}_1^{\alpha,\beta}$, for $N=3$. Even for larger values of $N$, considering samples of size $\lvert \mathscr{S} \rvert=500$, or $\lvert \mathscr{S} \rvert=2500$ produced numerically consistent results (as explained later in this section).

\begin{figure}[h!]
\begin{center}
\includegraphics[width=.9\textwidth]{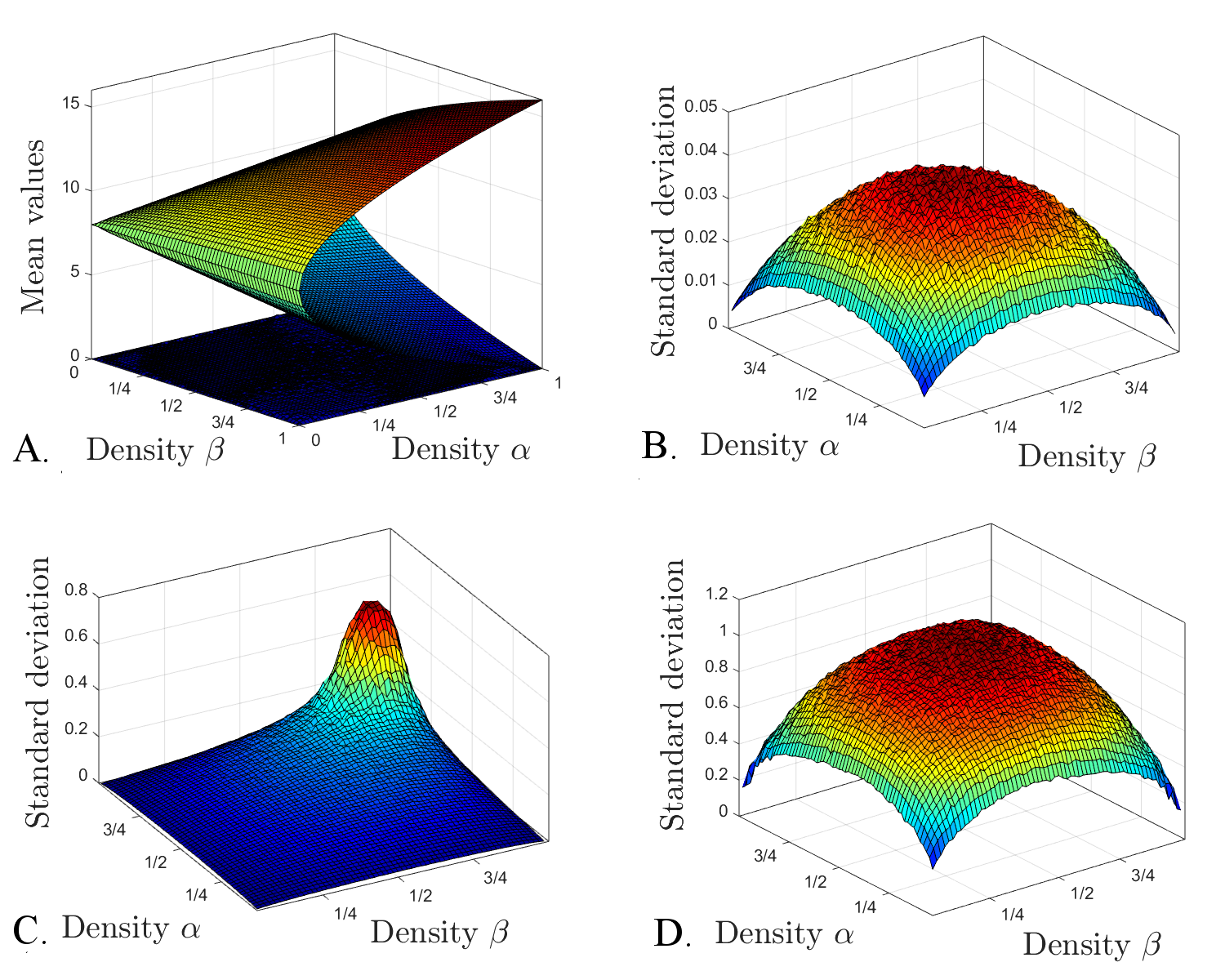}
\end{center}
\caption{\emph{ \footnotesize  {\bf Mean and standard deviation of eigenvalue magnitudes for $N=8$}, estimated numerically for each pair of densities $(\alpha,\beta)$ by considering a random sample of $2500$ matrices ${\bf T}$. {\bf A.} The mean eigenvalue real parts are represented as surfaces with respect to  $(\alpha,\beta)$. The two top surfaces fit very closely the expressions $N \pm N\sqrt{\alpha \beta}$; the other surfaces are all close to zero. For each of the first ({\bf B}), second ({\bf C}) and third ({\bf D}) leading eigenvalues (in magnitude), we represent the corresponding standard deviation as a surface with respect to $(\alpha,\beta)$.}}
\label{eigenvalues_8}
\end{figure}

\subsection{Numerical estimates of eigenvalue distributions}
\label{variance}

There are a few contexts in the literature on eigenspectra of random graphs that relate to our problem. The eigenspectrum of the adjacency matrix of a network with communities is known to have leading eigenvalues that are well separated from the rest of the spectrum~\cite{chauhan2009spectral}.

A result more qualitatively related to our question is due to Juh\'{a}sz~\cite{juhasz1990characteristic}. Viewed in the reference's general framework, the adjacency matrix ${\bf T}$ is a block matrix with (weighted) density matrix $\ds {\bf D}= \left[ \begin{array}{cc} 1 & \alpha \\ \beta & 1  \end{array} \right]$, whose eigenvalues are $\ds \mu_{1,2}=1 \pm \sqrt{\alpha \beta}$. According to the main theorem in the referenced paper, ${\bf T}$ has two eigenvalues $\lambda_{1,2}$ that are large (of order $N$) in magnitude, and the other eigenvalues close to zero. More precisely, $\ds \lambda_{1,2}=N \pm N \sqrt{\alpha \beta} + o(N^{1/2+\epsilon})$ in probability, while the other eigenvalues are of order $o(N^{1/2+\epsilon})$ in probability (for any $\epsilon > 0$).

A first thought is that $N \pm N \sqrt{\alpha \beta}$ may provide in our case the exact formal expressions for the means $E(\lvert \lambda_{1,2} \rvert)$  in terms of the densities $\alpha$ and $\beta$. The formulas look particularly promising, since they seem to naturally extend the boundary expressions obtained in the two lemmas (for $\alpha \in \{ 0,1 \}$ or $\beta \in \{ 0,1 \}$), and since they match tightly our numerical results (as shown in Figure~\ref{eigenvalues3}a and ~\ref{eigenvalues_8}a). Simple direct computations of the spectra for $N=3,4$ immediately reveal, however, that the formulas $\ds N \pm N \sqrt{\alpha \beta}$ do not give the exact means for the leading eigenvalue magnitudes -- although this may only be the case for finite sizes $N$, and the estimates may be in fact improving with increasing size, and may become exact in the limit $N \to \infty$. An interesting question to be addressed is that of understanding not only the shape of the leading eigenvalue distributions, but also the source of the error terms in their means compared to $\ds N \pm N \sqrt{\alpha \beta}$, and their own behavior with respect to the size $N$.

For the rest of the section, we gain a numerical insight, for size up to $N=20$, and provide a few numerically-based conjectures on the behavior of the spectrum as the size increases. In the technical Section~\ref{perturbation} we back up analytically some of the conjectures speculated in this section, based on our simulations.\\

\noindent Figure \ref{eigenvalues3} illustrates the standard deviation of ${\cal L}_1^{\alpha,\beta}$ as a function of the densities. For each pair $(\alpha,\beta)$ we computed the standard deviation of ${\cal L}_1^{\alpha,\beta}$ over all configurations in ${\cal D}^{\alpha,\beta}$ (Figure \ref{eigenvalues3}b), as well as over a random sample of $2500$ representatives for ${\bf T}$. Figures~\ref{eigenvalues_8}b,c and d show similar results for $N=8$; for each pair $(\alpha,\beta)$, we used $2500$ samples for ${\bf T}$ to estimate numerically the standard deviations of ${\cal L}_1^{\alpha,\beta}$, ${\cal L}_2^{\alpha,\beta}$ and ${\cal L}_3^{\alpha,\beta}$. In all cases, the surfaces decrease towards the edges, illustrating the narrowing of the corresponding distributions when $(\alpha,\beta)$ gets closer to the boundary of the unit square. We would like to point out the possible confound that the numerical scheme may be introducing by considering the same cardinality ($2500$) for sampling the larger distributions in the center, as well as the slimmer distributions near the boundary (i.e., the underestimation due to sampling may be more pronounced around the center of the surface than towards the boundary).

\begin{figure}
\begin{center}
\includegraphics[width=.9\textwidth]{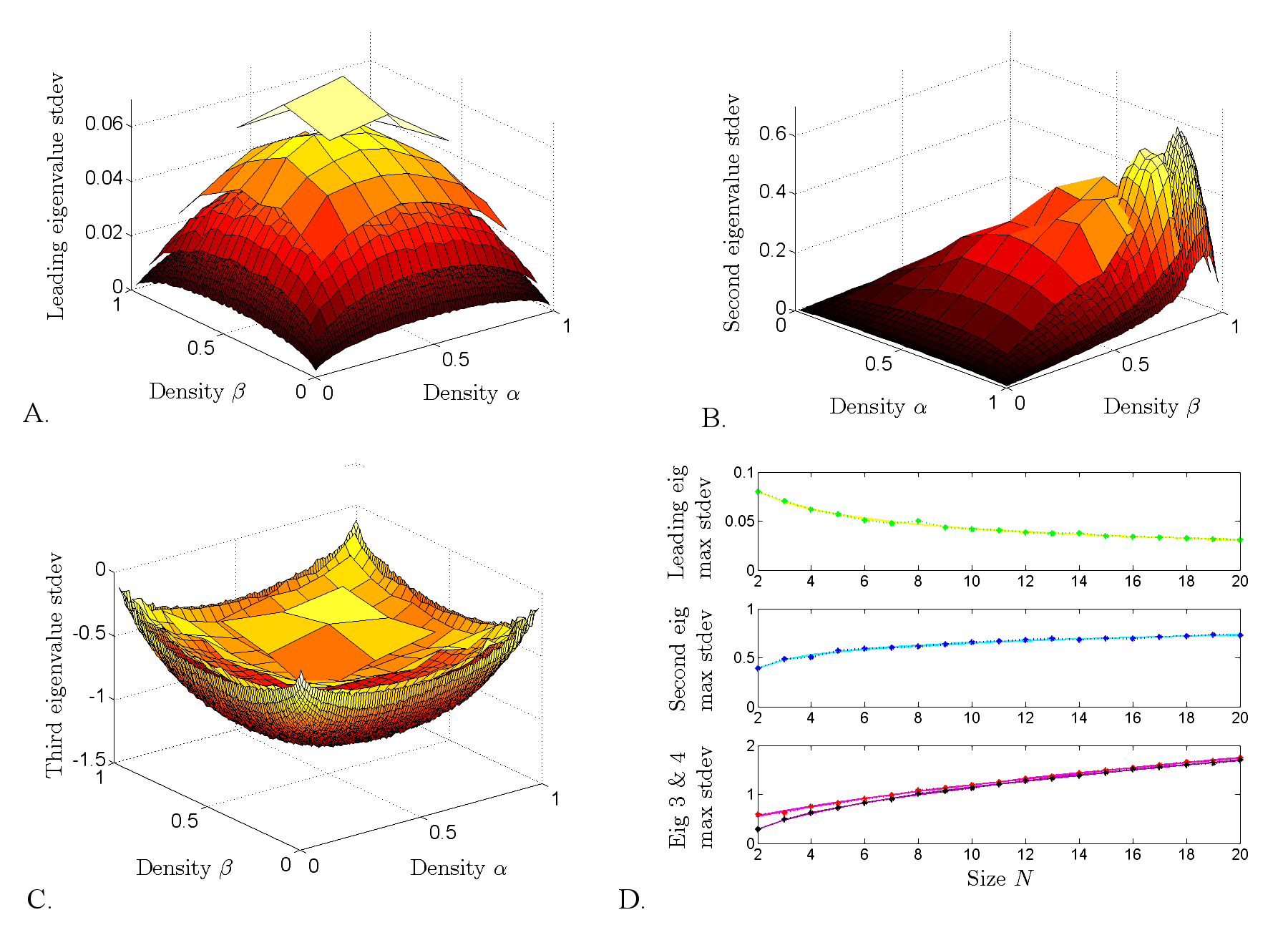}
\end{center}
\caption{ \emph{ \footnotesize {\bf Illustration of the evolution of the standard deviation of ${\cal L}_j^{\alpha,\beta}$, when increasing the network size $N$.} {\bf A.} Each surface represents the standard deviation of ${\cal L}_1^{\alpha,\beta}$ with respect to $(\alpha,\beta)$, for a different size $N$; from top to bottom: $N=2$, $N=3$, $N=5$ and $N=10$. {\bf B.} Each surface represents the standard deviation of ${\cal L}_2^{\alpha,\beta}$ for a different size $N$; from lowest to highest: $N=4$, $N=6$ and $N=10$. {\bf C.} Each surface represents the standard deviation of ${\cal L}_3^{\alpha,\beta}$, for a different size $N$; from top to bottom: $N=2$, $N=3$, $N=5$ and $N=10$. In order to make all surfaces visible, the figure is vertically flipped (we show minus the standard deviation). {\bf D.} The dotted plots show how the global maximum value of each surface evolves when increasing the size up to $N=20$. For each curve, we used a Levenberg-Marquardt algorithm to determine the best functional fit, shown as a solid line (in some cases the solid line is hard to see, because of its almost perfect overlap with the simulation data). Top: the maximum of ${\cal L}_1^{\alpha,\beta}$ decreases with $N$ (dotted green curve), as $\sim N^{-0.47}$ (yellow solid curve), with residuals norm $\epsilon$ = 0.0048. Middle:  the maximum of ${\cal L}_2^{\alpha,\beta}$ increases with $N$ (dotted blue curve), as $\sim \log(N)$ (cyan solid curve), with residuals norm $\epsilon$= 0.0013. Bottom: the maxima of ${\cal L}_3^{\alpha,\beta}$ (dotted red curve) and ${\cal L}_4^{\alpha,\beta}$ (dotted black curve)  increase as $\sim N^{0.59}$ (solid pink) and $\sim N^{0.51}$ (solid purple), with residual norms $\epsilon$=0.0034 and 0.0014 respectively. The estimates for {\bf A}, {\bf B} and {\bf C} are based on samples of size $2500$. The estimates for {\bf D} are based on samples of size $400$.}}
\label{stdev}
\end{figure}

For a fixed $N$, the distribution ${\cal L}_j^{\alpha,\beta}$  for each eigenvalue $\lambda_j$ is clearly largest at intermediate values of $\alpha$ and $\beta$. Following the same logic (``higher cardinality likely produces higher variance''), one would expect standard deviations to increase when the size $N$ is increased (recall that $\ds \lvert {\cal D}^{\alpha,\beta} \rvert = C_{N^2}^{\alpha N^2} C_{N^2}^{\beta N^2}$, which increases factorially with $N$). Juh\'{a}sz' estimate goes along the same lines, claiming an almost everywhere correction term of magnitude $o(N^{1/2+\epsilon})$, which increases with $N$. This means that there are almost no outliers out of the Juh\'{a}sz range, even though the  ``spread'' of each ${\cal L}_j^{\alpha,\beta}$ remains quite large (of order $o(N)$, as discussed in Section~\ref{perturbation}). 

In Figure~\ref{stdev}, we illustrate specifically the outcome of our numerical simulations of how the standard deviations behave with increasing $N$ (with approximation algorithms based on sample distributions). In Figures \ref{stdev}a,b,c we show, for $2 \leq N \leq 20$, the standard deviations for the three leading eigenvalues, each represented as a surface with respect to density pairs $(\alpha,\beta)$. Figure \ref{stdev}d tracks the behavior of the maximum of the surface corresponding to each of the first four eigenvalues,  over the unit $(\alpha,\beta)$ square. Our estimates suggest that, for $j=3,4$, the standard deviations of ${\cal L}_j^{\alpha,\beta}$ increase as a power function of $N$ (with the power $\sim1/2$). This is not surprising in light of the existing results already described. However, interestingly, the simulations suggest a decreasing power rule $\sim N^{-1/2}$ for the standard deviation of ${\cal L}_1^{\alpha,\beta}$, and a logarithmic increase for the standard deviation of ${\cal L}_2^{\alpha,\beta}$, implying that, for the two large eigenvalues, Juh\'{a}sz' result can be greatly refined in terms of standard deviations. This is a useful fact to investigate, since narrowness of the distributions ${\cal L}_{1,2}^{\alpha,\beta}$ with $N$ insures better separation between the leading eigenvalues and the rest of the spectrum, and subsequently more ``recognizable'' modularity properties (as discussed later in Section~\ref{general}). As mentioned before, this feature can become quite important when the graph operates as a functional network,e.g., as a brain feedback circuit.\\

\noindent We summarize our initial theoretical and numerical observations in the case of two connected cliques in the form of a conjecture, which remains open to a more rigorous investigation:

\begin{conj}
In the case of fully-connected modules $\gamma=1$ (i.e., ${\bf S}={\bf R}={\bf M}$), the spectrum of the matrix ${\bf T}$ varies with respect to the inter-modular densities $\alpha$ and $\beta$ of the blocks ${\bf A}$ and ${\bf B}$ as follows:

\begin{enumerate}[(i)]

\item For $(\alpha,\beta) \neq (1,1)$, the spectrum has two eigenvalues $\lambda_1$ and $\lambda_2$ whose mean magnitudes are large, while the other $2N-2$ have small mean magnitudes (close to zero). As $(\alpha,\beta) \to (1,1)$, the second large eigenvalue $\lambda_2 \to 0$ as well.

\item For each size $N$ and each density pair $(\alpha,\beta)$, the mean real parts of the two leading eigenvalues (over all adjacency configurations corresponding to $(\alpha,\beta)$), are given approximately by $N \pm N \sqrt{\alpha \beta}$, with error terms approaching zero as $N \to \infty$.

\item For any size $N$, the standard deviation of each eigenvalue's real part is a ``unimodal'' surface, with a point of maximum in the open square $(0,1)^2$, and which is zero when $\alpha \in \{ 0,1 \}$ or $\beta \in \{ 0,1 \}$.

\item For the leading eigenvalue $\lambda_1$, the standard deviations for all $(\alpha, \beta) \in [0,1]$ are very small. Moreover, the standard deviation of ${\cal L}_1^{\alpha,\beta}$ decreases monotonically with $N$, for each fixed pair $(\alpha,\beta)$.  The maximum attainable standard deviation of ${\cal L}_1^{\alpha,\beta}$ over $(\alpha, \beta) \in [0,1]$ decreases approximately as $N^{-1/2}$. (Note: This transcends qualitatively the corresponding Juh\'{a}sz estimate.)

\item For the second eigenvalue $\lambda_2$, the maximum attainable standard deviation of ${\cal L}_2^{\alpha,\beta}$ over $(\alpha, \beta) \in [0,1]$ increases logarithmically with $N$. (Note: This transcends quantitatively the corresponding Juh\'{a}sz estimate.)

\item For the rest of the eigenvalues $\lambda_j$, $j \geq 3$, the maximum attainable standard deviation of ${\cal L}_j^{\alpha,\beta}$ over $(\alpha, \beta) \in [0,1]$ increases approximately as $N^{1/2}$. (Note: This is the same as the rate of the almost everywhere error term previously obtained by Juh\'{a}sz.)

\end{enumerate}

\end{conj}

\noindent {\bf Remark.} We are in particular interested in understanding the robustness of the leading eigenvalues to changes in configuration, once the densities have been fixed. First, one might suspect that this robustness is due to a large extent to the existence of the two fully-connected cliques in our graph. In Appendix B, we investigate how results change when we relax the fully-connectedness condition. Second, recall that we are ultimately interested in whether robust features in the adjacency spectrum translate into robustness in dynamics (if we consider the corresponding network of coupled oscillators). In our follow-up paper (briefly previewed in Section ~\ref{adj_dynamics} and in Appendix A), we further discuss this aspect, and the potential connections between adjacency and dynamics classes.


\subsection{Estimates using perturbation theory}
\label{perturbation}

\noindent {\bf Notation.} Throughout this section, ${\bf M}$ will denote the $N \times N$ matrix with all entries equal to $1$, $\text{\large{\bf 1}}$ will denote the $N \times 1$ column vector with all entries $1$, and $\varphi$ will denote the function that computes the sum of all entries, for any arbitrary size matrix.\\ 

\noindent The adjacency matrix ${\bf T}$ for our graph is of the form : $\left[ \begin{array}{c|c} {\bf M} & {\bf A} \\ \cline{1-2} {\bf B} & {\bf M}  \end{array} \right]$, where $\varphi({\bf A})=\alpha N^2$ and $\varphi({\bf B})=\beta N^2$. At the start of Section~\ref{adjacency}, we have found the spectrum of ${\bf T}$ when $(\alpha,\beta)$ is on the boundary of the unit square. The spectrum is also easy to find for the matrix ${\bf C}=\left[ \begin{array}{c|c} {\bf M} & \alpha {\bf M} \\ \cline{1-2} \beta {\bf M} & {\bf M}  \end{array} \right]$, which is a non-binary matrix which ``averages out'' all configurations ${\bf T}$ for a fixed pair $(\alpha,\beta)$.

\begin{lemma} The matrix ${\bf C}=\left[ \begin{array}{c|c} {\bf M} & \alpha {\bf M} \\ \cline{1-2} \beta {\bf M} & {\bf M}  \end{array} \right]$ has eigenvalues:
\begin{itemize}
\item $\lambda_1=N + N\sqrt{\alpha \beta}$, with corresponding eigenvector ${\bf u}_1= \left[ \begin{array}{c} \sqrt{a} \text{\large{\bf 1}} \\ \sqrt{b} \text{\large{\bf 1}} \end{array} \right]$
\item $\lambda_2=N - N\sqrt{\alpha \beta}$, with corresponding eigenvector ${\bf u}_2= \left[ \begin{array}{c} \sqrt{a} \text{\large{\bf 1}} \\ -\sqrt{b} \text{\large{\bf 1}} \end{array} \right]$
\item $\lambda_3=\hdots = \lambda_{2N}=0$, with corresponding eigenspace spanned by the vectors ${\bf u}_k=\left[ \begin{array}{c} {\bf t}_k \\ {\bf s}_k \end{array} \right]$, where ${\bf t}_k$ and ${\bf s}_k$ are $N \times 1$ column vectors with $\varphi({\bf t}_k)=\varphi({\bf s}_k)=0$, for $k \geq 3$.
\end{itemize}
\label{lemma1}
\end{lemma}

\proof{ The proof is direct, and will be omitted.}

\noindent We are interested in the spectrum of the matrix ${\bf T}$, which we write in the form ${\bf T}={\bf C} + {\bf Z}$, where the error term ${\bf Z} = \left[ \begin{array}{c|c} {\bf 0} & {\bf U} \\ \cline{1-2} {\bf V} & {\bf 0}  \end{array} \right]$ has $\varphi({\bf U})=\varphi({\bf V})=0$. More generally, we consider the matrix family ${\bf T}_{\varepsilon} = {\bf C} + \varepsilon {\bf Z}$ (so that ${\bf T}={\bf T}_\varepsilon$, for $\varepsilon=1$). Notice that, with this notation, ${\bf T}_{\varepsilon}$ is a perturbation of ${\bf T}$ of order $O(\varepsilon)$. The leading eigenvalue (and its corresponding eigenvector) of ${\bf T}$ can then be thought of as a perturbation of the original leading eigenvalue $\lambda_1$ of ${\bf T}$ (with corresponding original eigenvector ${\bf u}_1$). Using a perturbation theory approach, we can compute the first correction term in the expansion of $\lambda_1$:

\begin{equation}
{\bf T}_\varepsilon ({\bf u}_1 + \varepsilon {\bf x}) = (\lambda_1+\varepsilon \mu_1)({\bf u}_1 + \varepsilon {\bf x})
\label{perturb}
\end{equation}

\noindent We may assume without loss of generality that $\varepsilon$ is arbitrarily small (since the magnitude of ${\bf Z}$ is arbitrary), and that the direction ${\bf x}$ of the eigenvector perturbation is perpendicular to ${\bf u}_1$. Expanding with respect to $\varepsilon$ and identifying the coefficients of $\varepsilon$, we get:

$${\bf Z u}_1 + {\bf Cx} = \lambda_1{\bf x} + \mu_1 {\bf u}_1$$

\noindent In the basis $({\bf u}_k)_{k=\overline{1,2N}}$ of eigenvectors of ${\bf C}$, one can write ${\bf x}=\sum x_k {\bf u}_k$, and the matrix ${\bf Z}$ as $(z_{jk})_{i,j=\overline{1,2N}}$, so that ${\bf Z u}_j = \sum z_{jk} {\bf u}_k$. Then our equation becomes:

$$\sum x_k \lambda_k {\bf u}_k + \sum z_{1k} {\bf u}_k = \lambda_1 \sum x_k {\bf u}_k + \mu_1 {\bf u}_1$$

\noindent Solving in components, this gives us:

\begin{itemize}

\item $\mu_1 = z_{11}$, for $k=1$

\item $\ds x_k = \frac{z_{1k}}{\lambda_1-\lambda_k}$, for $k \geq 2$.

\end{itemize}

\noindent Given the form of the eigenvectors in the basis $({\bf u}_k)_{k=\overline{1,2N}}$, we easily can calculate some of the $z_{jk}$s that are most useful to continue our computation. For example, on one hand:

$${\bf Zu}_1 = \left[ \begin{array}{c|c} {\bf 0} & {\bf U} \\ \cline{1-2} {\bf V} & {\bf 0}  \end{array} \right] \left[ \begin{array}{c} \sqrt{a} \text{\large{\bf 1}} \\ \sqrt{b} \text{\large{\bf 1}} \end{array} \right] = \left[ \begin{array}{c} \sqrt{b} {\bf U}\text{\large{\bf 1}} \\ \sqrt{a} {\bf V} \text{\large{\bf 1}} \end{array} \right]$$

\noindent and on the other hand, in components:

$${\bf Zu}_1 = z_{11} \left[ \begin{array}{c} \sqrt{a} \text{\large{\bf 1}} \\ \sqrt{b} \text{\large{\bf 1}} \end{array} \right] + z_{12} \left[ \begin{array}{c} \sqrt{a} \text{\large{\bf 1}} \\ -\sqrt{b} \text{\large{\bf 1}} \end{array} \right] + \sum_{k \geq 3} z_{1k} \left[ \begin{array}{c} {\bf t}_k \\ {\bf s}_k \end{array} \right]$$

\noindent Recall that $\varphi({\bf t}_k)=\varphi({\bf s}_k)=0$, for all $k \geq 3$. applying the operator $\varphi$ separately over the first the top and bottom $N$ entries, we get, respectively:

\begin{eqnarray}
N\sqrt{a}(z_{11} + z_{12}) &=& \sum_{k \geq 3} z_{1k}\varphi({\bf t}_k) = 0 \nonumber \\
N\sqrt{b}(z_{11} - z_{12}) &=& \sum_{k \geq 3} z_{1k}\varphi({\bf s}_k) = 0 \nonumber
\end{eqnarray}

\noindent This implies that $z_{11}=z_{12}=0$, and subsequently $\mu_1=0$. Hence the $O(\varepsilon)$ approximation of the leading eigenvalue of ${\bf T}_{\varepsilon}$ is $\lambda_1 = N+N\sqrt{\alpha \beta}$ (the leading eigenvalue of ${\bf C}$). One can continue in a similar fashion to obtain higher order approximations. To get the second correction term, we rewrite Eq.~\eqref{perturb} to include higher order terms:

\begin{equation}
{\bf T}_\varepsilon ({\bf u}_1 + \varepsilon {\bf x} + \varepsilon^2 {\bf y}) = (\lambda_1+\varepsilon \mu_1+\varepsilon^2 \nu_1)({\bf u}_1 + \varepsilon {\bf x} + \varepsilon^2 {\bf y})
\label{perturb2}
\end{equation}

\noindent As before, we can assume without loss of generality that the direction of ${\bf y}$ is perpendicular to that of ${\bf u}_1$. Identifying the coefficients of $\varepsilon^3$ and $\varepsilon^4$, and using the fact that $\mu_1=0$, we get two more equations, which can be used to completely determine $\nu_1$ and ${\bf y}$:

\begin{eqnarray}
{\bf Cy} &+& {\bf Zx} = \lambda_1 {\bf y} + \nu_1{\bf u}_1 \label{order2}\\
{\bf Zy} &=& \nu_1{\bf x} \nonumber
\end{eqnarray}

\noindent Writing equation~\eqref{order2} in components, we have:

$$\sum_{k} \lambda_k {\bf y}_k {\bf u}_k + \sum_{k,l} {\bf x}_k z_{kl} {\bf u}_l = \lambda_1 \sum_k {\bf y}_k {\bf u}_k + \nu_1 {\bf u}_1$$

\noindent Projecting in the direction of ${\bf u}_1$, replacing $\ds x_k = \frac{z_{1k}}{\lambda_1-\lambda_k}$, for all $k \geq 2$, and also recalling that $z_{11}=0$, we have that:

\begin{equation}
\nu_1 = \sum_k {\bf x}_k z_{k1} = \sum_{k \geq 2} \frac{z_{1k}z_{k1}}{\lambda_1-\lambda_k}
\end{equation}

\noindent We additionally know that $z_{12}=0$ and that $\lambda_k=0$, for $k \geq 3$. Hence:

\begin{equation}
\nu_1 = \sum_{k \geq 3} \frac{z_{1k}z_{k1}}{\lambda_1} = \frac{1}{N+N\sqrt{\alpha \beta}} \sum_{k \geq 3} z_{1k}z_{k1}
\end{equation}

\noindent But $\sum_{k \geq 3} z_{1k}z_{k1} = \sum_{k \geq 1} z_{1k}z_{k1}$ is in fact nothing but the first component of the matrix ${\bf Z}^2$, written in the basis $({\bf u}_k)_{k=\overline{1,2N}}$. In other words, if we write in components $\ds {\bf Z^2u}_1 = A_1{\bf u}_1+A_2{\bf u}_2 + \sum_{k \geq 3} A_k{\bf u_k}$, then $\sum_{k \geq 1} z_{1k}z_{k1} = A_1$. To calculate $A_1$, we can use the fact that $\langle {\bf u}_1,{\bf u}_k \rangle = \langle {\bf u}_2,{\bf u}_k \rangle=0$, for all $k \geq 3$, and calculate:

\begin{eqnarray}
\langle {\bf u}_1, {\bf Z}^2{\bf u}_1 \rangle = A_1 \| {\bf u}_1 \| + A_2 \langle {\bf u}_1, {\bf u}_2 \rangle = N(a+b)A_1 + N(a-b)A_2 \nonumber \\
\langle {\bf u}_1, {\bf Z}^2{\bf u}_1 \rangle = A_1 \langle {\bf u}_1, {\bf u}_2 \rangle + A_2 \| {\bf u}_2 \| = N(a-b)A_1 + N(a+b)A_2 \nonumber
\end{eqnarray}

\noindent On the other hand, ${\bf Z}^2 = \left[ \begin{array}{c|c} {\bf UV} & {\bf 0} \\ \cline{1-2} {\bf 0} & {\bf VU}  \end{array} \right]$, so that:

\begin{eqnarray}
\langle {\bf u}_1, {\bf Z}^2{\bf u}_1 \rangle = \alpha \text{\large{\bf 1}}^T {\bf UV} \text{\large{\bf 1}} + \beta \text{\large{\bf 1}}^T {\bf VU} \text{\large{\bf 1}} \nonumber \\
\langle {\bf u}_2, {\bf Z}^2{\bf u}_1 \rangle = \alpha \text{\large{\bf 1}}^T {\bf UV} \text{\large{\bf 1}} - \beta \text{\large{\bf 1}}^T {\bf VU} \text{\large{\bf 1}} \nonumber
\end{eqnarray}

\noindent Combining the two, we get:

\begin{eqnarray}
N(A_1+A_2) = \text{\large{\bf 1}}^T {\bf UV} \text{\large{\bf 1}}  \nonumber \\
N(A_1-A_2) = \text{\large{\bf 1}}^T {\bf VU} \text{\large{\bf 1}} \nonumber
\end{eqnarray}

\noindent hence $\ds A_1 = \frac{1}{2N} \left( \text{\large{\bf 1}}^T {\bf UV} \text{\large{\bf 1}} + \text{\large{\bf 1}}^T {\bf VU} \text{\large{\bf 1}} \right)$.\\

\noindent In conclusion, we have the following:

\begin{prop} The leading eigenvalue of ${\bf T}_\varepsilon$ is

$$\lambda_1 = N+N\sqrt{\alpha \beta} + \varepsilon^2 \frac{1}{2N} \; \frac{1}{N+N\sqrt{\alpha \beta}} \left( \text{\large{\bf 1}}^T {\bf UV} \text{\large{\bf 1}} + \text{\large{\bf 1}}^T {\bf VU} \text{\large{\bf 1}} \right) + O(\varepsilon^3)$$
\label{perturb}
\end{prop}

\vspace{4mm}
\noindent This gives us, in particular, a better approximation of the leading eigenvalue of ${\bf T}={\bf T}_{\varepsilon=1}$, using two correction terms. Note that one can easily obtain bounds of order $N^3$ for the term $\varphi({\bf UV+VU})= \text{\large{\bf 1}}^T {\bf UV} \text{\large{\bf 1}} + \text{\large{\bf 1}}^T {\bf VU} \text{\large{\bf 1}}$.\\

\begin{lemma} Consider two $N \times N$  binary matrices ${\bf A}$ and ${\bf B}$ with densities of ones $\alpha$ and respectively $\beta$, that is $\varphi({\bf A})=\alpha N^2$ and $\varphi({\bf B})=\beta N^2$. Then: $\ds \varphi({\bf AB}) \leq N^3 \sqrt{\alpha \beta}$
\label{lemma_variability}
\end{lemma}

\proof{ Notice that 

$$\varphi({\bf AB}) = [{\bf A}^1... {\bf A}^N] \cdot \left[ \begin{array}{c} {\bf B}_1\\ \vdots \\ {\bf B}_N \end{array} \right]$$

\noindent where ${\bf A}^i$ is the sum of the elements in the $i$-th column of ${\bf A}$, and ${\bf B}_i$ is the sum of the elements in the $i$-th row of ${\bf B}$, hence $\sum{\bf A}^i=\alpha N^2$ and $\sum{\bf B}_i=\beta N^2$. Using the Cauchy-Schwartz inequality, we can see that

$$\varphi({\bf AB}) = \sum {\bf A}^i {\bf B}_i \leq \sqrt{\sum ({\bf A}^i)^2 \cdot \sum ({\bf B}_i)^2}$$

\noindent Furthermore, each $\ds ({\bf A}^i)^2 = (\sum a_{i1})^2 \leq \sum a_{i1}^2 \cdot \sum 1 = N \sum a_{i1}$. Similarly, each $\ds {\bf B}_i \leq N \sum b_{1i}$, hence $\ds \sqrt{\sum ({\bf A}^i)^2 \cdot \sum ({\bf B}_i)^2} \leq \sqrt{N \varphi({\bf A}) \cdot N \varphi({\bf B})} = N^3 \sqrt{\alpha \beta}$.\\

\noindent In conclusion, $\ds \varphi({\bf AB}) \leq N^3 \alpha \beta$.
}

\begin{corol}With the existing notations, we have
$$-2N^3 \alpha \beta \leq \varphi({\bf UV}) \leq 2N^3 \sqrt{\alpha \beta} (1-\sqrt{\alpha \beta})$$
\end{corol}

\proof{We use the fact that ${\bf U}={\bf A}-\alpha{\bf M}$ and ${\bf V}={\bf B}-\beta{\bf M}$, to compute: $\varphi({\bf UV})=\varphi({\bf AB} - \alpha {\bf BM} - \beta {\bf AM} + \alpha \beta {\bf M}^2)=\varphi({\bf AB})-\alpha \beta N^3 - \alpha \beta N^3 + \alpha \beta N^3 = \varphi({\bf AB})-\alpha \beta N^3$.
}

This implies a first correction term of order $\ds \frac{1}{2N} \;\frac{1}{N+N\sqrt{\alpha \beta}} \cdot N^3$ for the eigenvalue $\lambda_1$ around $N+N\sqrt{\alpha \beta}$. This is not helpful if our aim is to narrow down the estimates as $N$ increases, since this bound increases like $N$ with the size, presenting similar problems with Juhazs' estimates for the probabilistic case, and failing to explain our numerical conjectures. However, let us notice that this estimate can't be qualitatively improved (in the sense of sharpening it to a lower order of $N$), since, for all $N$, one can always find outliers in the distribution ${\cal L}^{\alpha,\beta}_1$ at a distance $\sim N$ from $N+N\sqrt{\alpha \beta}$. An explanation that reconciles both observations, as well as Juh\'{a}sz' almost everywhere bounds,  is that these outliers are less representative as $N$ increases, causing the distribution to remain narrow, with a small \emph{ standard deviation} that decreases with $N$.\\

\begin{figure}[h!]
\begin{center}
\includegraphics[width=\textwidth]{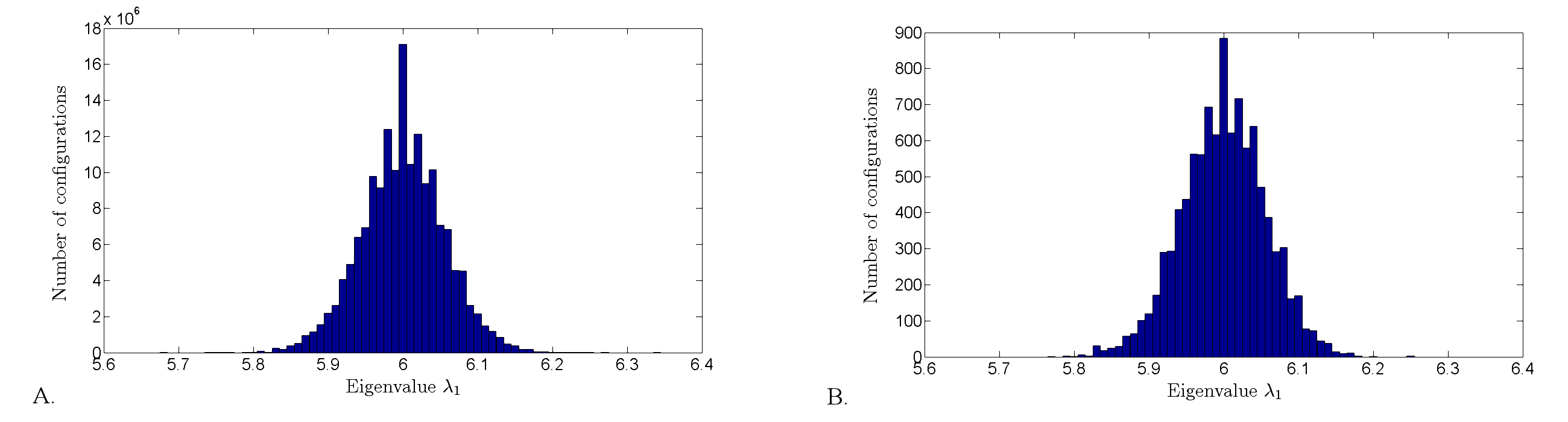}
\end{center}
\caption{ \emph{ \footnotesize {\bf Comparison between the distribution ${\cal L}^{\alpha,\beta}_1$ and a sample based distribution}. {\bf A.} Histogram of the distribution ${\cal L}^{\alpha,\beta}_1$, for $N=4$, $\alpha=8$, $\beta=8$y. {\bf B.} Histogram for a random subset of values in ${\cal L}^{\alpha,\beta}_1$, for $N=4$, $\alpha=8$, $\beta=8$, computed based on a sample of size $10^4$ configurations out of the total of.}}
\label{hist}
\end{figure}

\noindent These statements seem rather difficult to support, due on one hand to the difficulty of a direct analytic calculation of the standard deviation, and on the other hand to the potential inaccuracies in the numerical computations of the standard deviation based on fixed size sample distributions. Indeed, recall that the size of ${\cal L}^{\alpha,\beta}_1$ increases factorially with $N$, making it unrealistic to explore all configurations in this distribution. Hence any computationally tractable approach based on sample distributions can only increase the sample sizes with $N$ at a much slower rate than the rate at which the actual size of  ${\cal L}^{\alpha,\beta}_1$ increases, making these samples potentially  less and less reliable with larger sizes. For a brief illustration of the appropriateness of our sample-based computations, we compare in Figure~\ref{hist} the histogram of  ${\cal L}^{\alpha,\beta}_1$ (containing, for $N=4$ and $\alpha=\beta=8$, a total of $12,870^2$ configurations) with that produced by a sample of $100^2$ configurations.

Let us recall that one additional difficulty in calculating the standard deviation is the fact that the exact value of the mean of ${\cal L}^{\alpha,\beta}_1$ is not known, so a more basic task is to find a tight estimate for this mean. It is in this direction that an expression such as that obtained in Proposition~\ref{perturb} becomes directly useful. If one considers, for example, the first error term in the expansion in Proposition~\ref{perturb}, it is easy to show that, although the variability of $\varphi({\bf UV+VU})$ increases with $N$, the mean of this quantity over all configurations is zero.

\begin{lemma} Consider two $N \times N$  binary matrices ${\bf A}$ and ${\bf B}$ with densities $\alpha$ and respectively $\beta$. Then $E(\varphi({\bf AB}))=N^3 \alpha \beta$, where $E$ represents  the mean over all configurations in ${\cal D}^{\alpha,\beta}$ (i.e., with fixed $\varphi({\bf A})=\alpha$ and $\varphi({\bf B})=\beta$).
\label{lemma_mean}
\end{lemma}

\proof{With the previous notation, we have: $E({\bf A}^i)=\alpha N$ and $E({\bf B}_i)=\beta N$, for all $1 \leq i \leq N$. Since the matrices ${\bf A}$ and ${\bf B}$ are independent, we can easily compute $E(\varphi({\bf AB})=N \cdot \alpha N \cdot \beta N = \alpha \beta N^3$.
}

\begin{corol}
The mean $E( \text{\large{\bf 1}}^T {\bf UV} \text{\large{\bf 1}} + \text{\large{\bf 1}}^T {\bf VU} \text{\large{\bf 1}})=0$.
\end{corol}

\noindent  The corollary follows directly from Lemma~\ref{lemma_mean}, and shows that the second correction term in the perturbation expansion of $\lambda_1$ is zero in mean. The computation can be continued to obtain higher order approximations, providing a heuristic understanding of what makes the mean $E({\lambda}_1$) remain close to $N+N\sqrt{\alpha \beta}$ for all values of $\alpha$, $\beta$ and $N$.


\section{Dependence of Laplacian spectrum of network size and edge densities}
\label{Laplacian}

For our oriented graph with adjacency matrix $\ds {\bf T}= \left[ \begin{array}{c|c} {\bf M} & {\bf A} \\ \cline{1-2} {\bf B} & {\bf M}  \end{array} \right]$, we consider the in node degree diagonal matrix ${\bf \Delta}$, with:

$${\bf \Delta}_{jj} = \varphi_j({\bf T}) \text{ for all } j =\overline{1,2N}$$

\noindent so that the corresponding Laplacian matrix is given by: $\ds {\bf L} = {\bf \Delta} - {\bf T}$.

The Laplacian eigenvalue spectrum has been used as a measure of system dynamics. For example, the algebraic connectivity, defined as the second smallest eigenvalue $\mu_{N-1}$ of the discrete Laplacian matrix, is known to play an important role on synchronization dynamics, network robustness, etc. In an effort to study the effect of interdependent topologies on the mutual synchronization of networks, Martin-Hernandez et al.~\cite{martin2013synchronization} focused on computing and approximating the algebraic connectivity of two interdependent networks, and on was showing that it experiences a phase transition upon the addition of a sufficient number of links among two interdependent networks. Here, we study the dependence of the Laplacian eigenvalues on the densities $(\alpha,\beta)$. 

\begin{figure}[h!]
\begin{center}
\includegraphics[width=.8\textwidth]{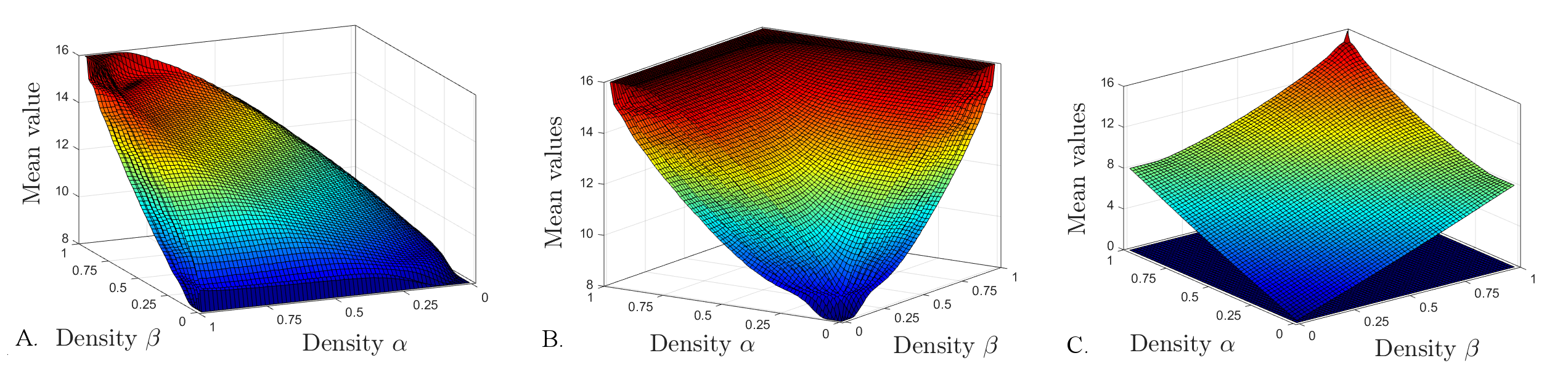}
\end{center}
\caption{ \emph{ \footnotesize  {\bf Mean of eigenvalue real parts for $N=8$}, estimated numerically for each pair of densities $(\alpha,\beta)$ by considering the same random sample of $2500$ adjacency configurations as in Figure~\ref{eigenvalues_8}a. Ordered by their magnitude: {\bf A.} the leading $N-1$ eigenvalues (1 through 6); {\bf B.} the following $N-1$ eigenvalues (7 through 14); {\bf C.} the two smallest eigenvalue (15 and 16). The smallest eigenvalue is zero (the Laplacian matrix is always rank degenerate).}}
\label{L_eigenvalues_8}
\end{figure}

\begin{figure}[h!]
\begin{center}
\includegraphics[width=.7\textwidth]{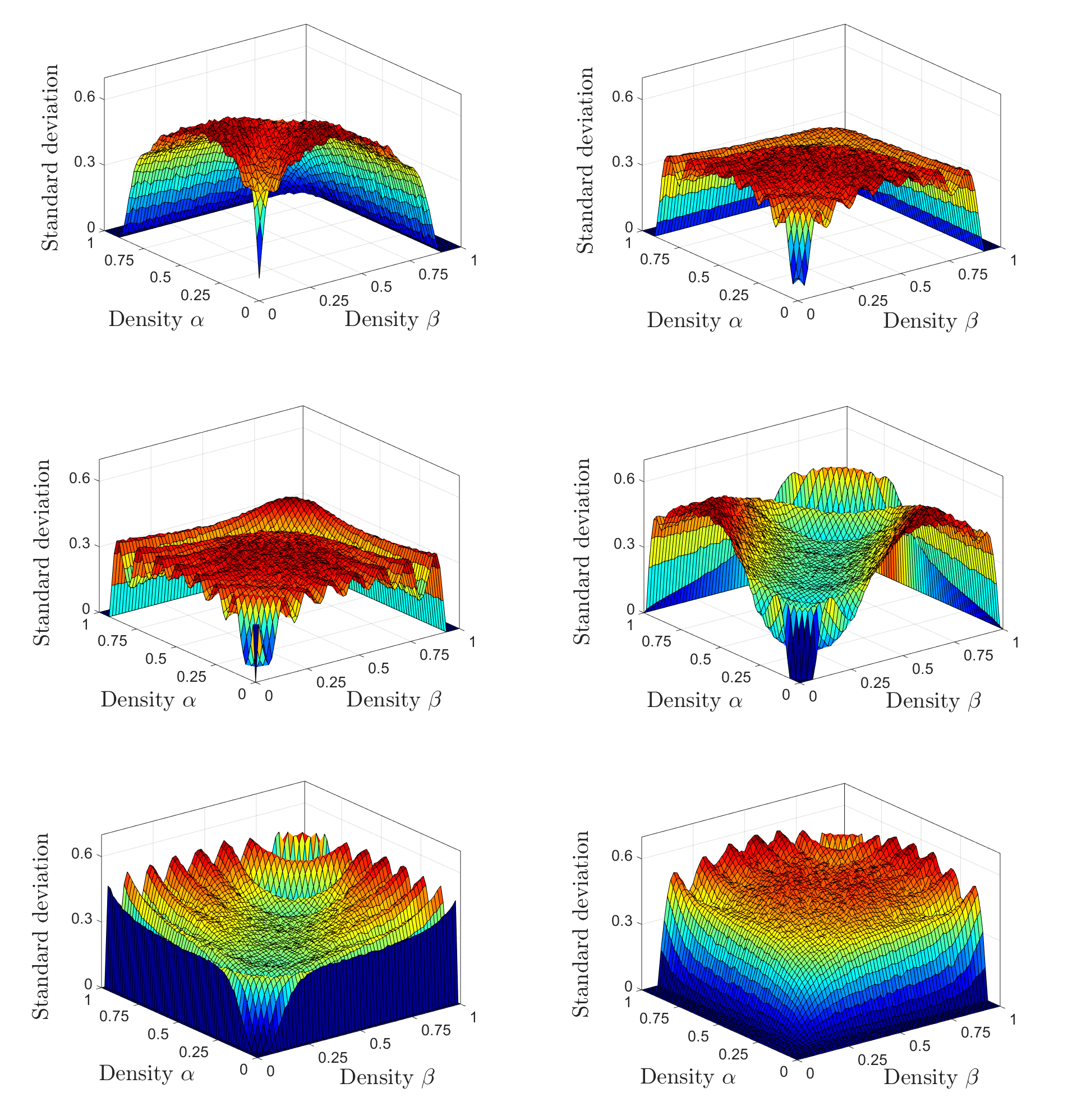}
\end{center}
\caption{ \emph{ \footnotesize  {\bf Standard deviations of eigenvalue real parts for $N=8$}. The panels represent, from top to bottom and left to right, the standard deviations for the eigenvalues 1, 2, 3, 7, 10, 14 (ordered by magnitudes).}}
\label{L_stds_8}
\end{figure}

Following the same numerical scheme as in Section~\ref{adjacency}, we computed the Laplacian eigenvalues for a sample of configurations, chosen randomly from the large distribution of all configurations corresponding to any fixed density pair $(\alpha,\beta)$. Based on this sample, we estimated, for each $(\alpha,\beta)$ the mean and standard deviation of the real part of the spectrum, as illustrated in Figures~\ref{L_eigenvalues_8} and~\ref{L_stds_8} for $N=8$.

\begin{figure}[h!]
\begin{center}
\includegraphics[width=.7\textwidth]{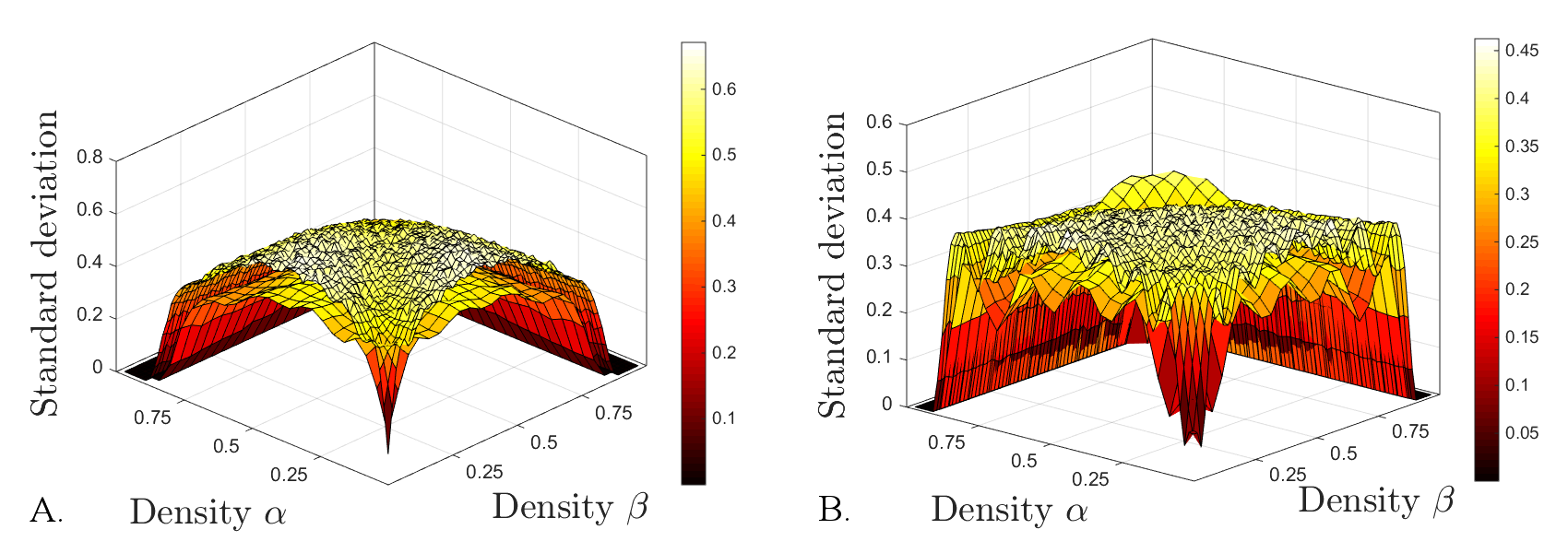}
\end{center}
\caption{ \emph{ \footnotesize {\bf Illustration of the evolution of the standard deviation of the Laplacian eigenvalue real parts, when increasing the network size $N$.} {\bf A.} Surfaces for the first eigenvalue, computed for $N=5$ (lower surface) and $N=8$ (higher surface). {\bf B.} Surfaces for the second eigenvalue, for $N=5$ (lower surface) and $N=8$ (higher surface). }}
\label{compare}
\end{figure}

\begin{figure}[h!]
\begin{center}
\includegraphics[width=\textwidth]{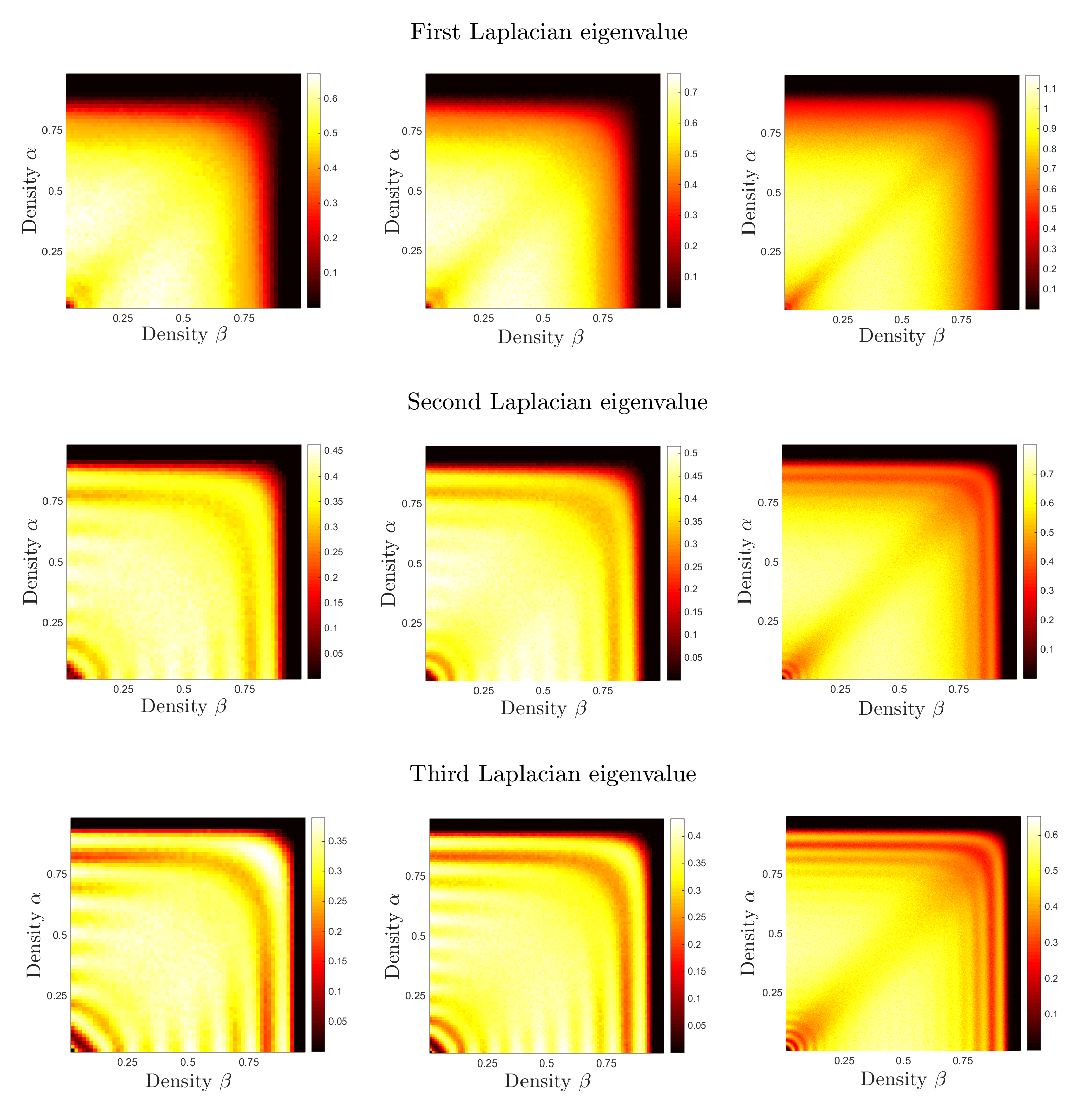}
\end{center}
\caption{ \emph{ \footnotesize {\bf Illustration of the evolution of the standard deviation of the Laplacian eigenvalue real parts, when increasing the network size $N$.}  The surfaces for the first second and third eigenvalues are shown top to bottom as pcolor plots, for $N=5$ (left), $N=10$ (center) and $N=20$ (right).}}
\label{pcolor}
\end{figure}

The behavior of the standard deviations for the real parts of the Laplacian eigenvalues with respect to the density pair $(\alpha,\beta)$ is very different than that of the standard deviations for the adjacency spectrum. While the adjacency standard deviation surfaces were unimodal on the domain $[0,1]^2$, decreasing from a central peak towards the boundary, in the case of the Laplacian, the surfaces are rippled (Figure~\ref{L_stds_8}), with the amplitude and distribution of the ripples depending on a variety of factors (as illustrated in Figure ~\ref{pcolor}, and discussed below).

Such variability in the standard deviation values makes it easier for the system to switch from robust regimes (with a narrow distribution of eigenvalues), to more scattered regimes (with a wider distribution of potential eigenvalues) by introducing a small change in the density $(\alpha,\beta)$. Scattered regimes are more sensitive to configuration, since wide changes in the Laplacian spectrum (and implicitly in Laplacian-driven dynamics) are accessible even under the same density pair by slightly altering the configuration. This could be in principle viewed as an adaptability feature that makes Laplacian-driven a desirable type of dynamics. 

\begin{figure}[h!]
\begin{center}
\includegraphics[width=.7\textwidth]{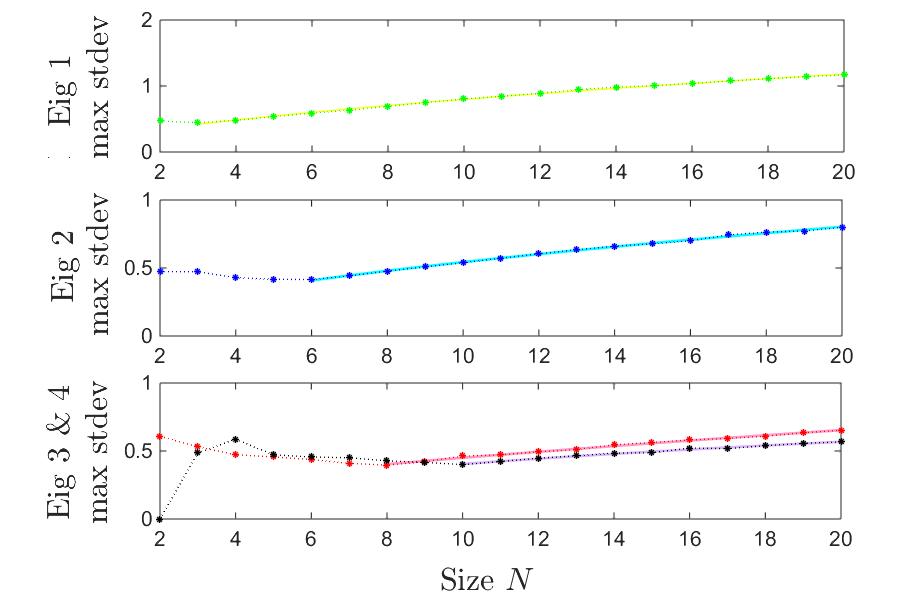}
\end{center}
\caption{ \emph{ \footnotesize {\bf Title.} The dotted plots show how the global maximum value of each surface evolves when increasing the size up to $N=20$. For each curve, we used a Levenberg-Marquardt algorithm to determine the best functional fit, shown as a solid line. Top: the maximum mean real part for the first eigenvalue increases with $N \geq 3$ (dotted green curve), as $\sim N^{0.42}$ (yellow solid curve), with residuals norm $\epsilon \sim 10^{-3}$. Middle:  the maximum for the second eigenvalue increases with $N \geq 6$ (dotted blue curve), as $\sim N^{0.52}$ (cyan solid curve), with residuals norm $\epsilon \sim 10^{-4}$. Bottom: the maxima for the third (dotted red curve) and fourth eigenvalues (dotted black curve)  increase as $\sim N^{0.69}$ (solid pink) and $\sim N^{0.33}$ (solid purple) for $N \geq 8$ and $N \geq 10$ respectively, with residual norms $\epsilon ~\sim 10^{-4}$. The estimates are based on samples of size $500$.}}
\label{fitting}
\end{figure}

However, the emergent robustness observed in the case of the adjacency leading eigenvalue (standard deviation of the real part decreasing with the size $N$) does not hold in the case of the leading Laplacian eigenvalue. In fact, the maximim standard deviations over the $(\alpha,\beta)$ domain seem to increase as powers of $N$ for all the eigenvalues in the Laplacian spectrum, after an initial transient phase for very small $N$ (see Figures ~\ref{compare},~\ref{pcolor} and~\ref{fitting}). As $N$ increases, the central regions of the surface, which raise with $N$, smoothen out and in the process push the ripples towards the borders.

If comparing the behavior of the two (adjacency and Laplacian) spectra when changing $(\alpha,\beta)$ and increasing $N$, once could say that the desirable feature of the adjacency model is robustness of the leading eigenvalue which \emph{ increases} with size, while the feature of the Laplacian model is swiftness between robust and loose regimes, which \emph{ degrades} with increasing size.

\section{Discussion}
\label{discussion}

\subsection{Comparison with random graphs approaches to modularity}
\label{general}

In this study, we have investigated, using analytical and numerical computations, the adjacency spectrum of an oriented graph, in which the nodes of two modules connect though fixed numbers of random edges within each module, as well as across modules. We concluded that, when fixing the number of both intra and inter-modular edges, the adjacency spectrum of the network remains in general sufficiently robust under particular edge configurations (geometries), suggesting that simple algorithms in such a system may also remain unaffected by constrained geometry changes.

There is a very large body of work addressing properties of random matrices~\cite{tao2012topics}, whose entries are drawn independently out of a given (typically normal) probability distribution. If, in addition, the matrix represents the adjacency of a random graph, so that each entry equals $1$ with a given probability, there are classical methods used when looking for properties of the the spectrum (e.g., spectral radius, or spectral density). Our model differs from most of these approaches in that it conserves the number of edges within/between modules, rather than fixing the independent probability of having an edge that connects two given nodes in the same/different modules. In our setup, the entries of the adjacency matrix are neither independent, nor identically distributed. However, while classical results (such as Wigner's semicircle law) require the entries to be identically distributed, various extensions have been worked out, for models which don't necessarily abide by these properties.

Consider, for example, the configuration model~\cite{farkas2001spectra}, whose spectral properties have been addressed by numerous studies. Since its edges are not statistically independent, a direct analytical approach is very difficult; existing results range from approximating the full spectrum~\cite{dorogovtsev2003spectra}, to formally deriving the expected values of the leading eigenvalue, but only in the large $N$ limit~\cite{chung2003spectra}. In a recent paper, Newman et al.~\cite{nadakuditi2013spectra} took an indirect approach: they considered instead a model with the same degree sequence as the configuration model, but in which the number of edges between any two nodes was drawn independently from a Poisson distribution. They then showed that the spectra of the two models agree in the large $N$ limit.

Below, we illustrate the same idea, by carrying out a large $N$ limit comparison between our model and its probabilistic counterpart, with independent, stochastic edges, considered by Nadakuditi and Newman~\cite{nadakuditi2012graph}. In the reference, the authors considered a stochastic, non-oriented network with two communities, and computed the ensemble-means for the two large eigenvalues of its symmetric adjacency matrix, in the large $N$ limit. The method involved first finding the eigenvalues of the modularity matrix, then showing that these are identical in the large $N$ limit to the eigenvalues of the adjacency matrix. Their asymptotic expressions $z_1$ and $z_2$ were computed in terms of $c_\text{in}=n p_\text{in}$ and $c_\text{out}=n p_\text{out}$ (where the notations in the original text are $n$ for the matrix size, $p_\text{in}$ for the probability of two nodes within a module to be directly connected, and $p_\text{out}$ for the probability of two nodes which are not in the same module to be directly connected). More precisely:

\begin{eqnarray*}
z_1 &=& \frac{1}{2} (c_\text{in}+c_\text{out})+1\\
z_2 &=& \frac{1}{2} (c_\text{in}-c_\text{out}) + \frac{c_\text{in}+c_\text{out}}{c_\text{in}-c_\text{out}}
\end{eqnarray*}

\noindent With our notation, $c_\text{in}=2N$, $c_\text{out}=2\alpha N$ and the adjacency matrix is symmetric ($\beta=\alpha$). Accounting for the presence of loops in our network (which were excluded in the Nadakuditi-Newman model), we get:

\begin{eqnarray*}
z_1 &=& N + \alpha N\\
z_2 &=& N (1-\alpha)+\frac{1+\alpha}{1-\alpha}-1
\end{eqnarray*}

\begin{figure}[h!]
\begin{center}
\includegraphics[width=.9\textwidth]{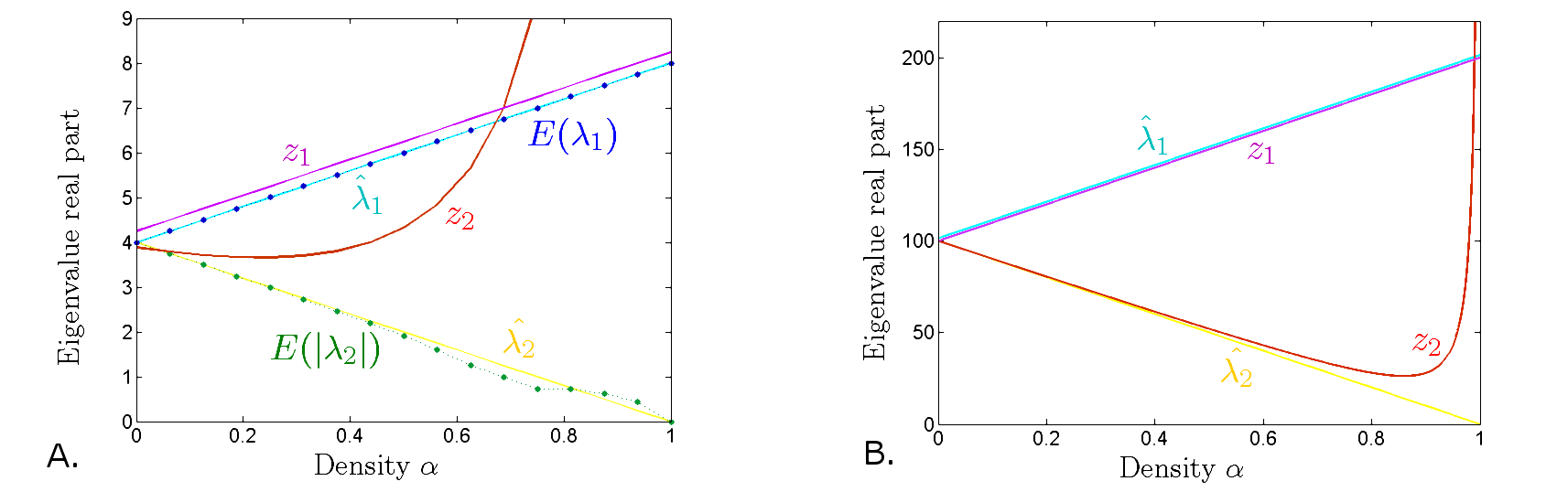}
\end{center}
\caption{\emph{\footnotesize {\bf Comparison between our results and those of Nadakuditi-Newman, in the case of a bimodular, non-oriented graph.} We compare the values of $z_{1,2}$ (solid curves in purple and brown, respectively) with the formal means $E({\lambda}_{1,2})$ (dotted curves in blue and green), and their close approximations $\ds N \pm \alpha N$ (solid curves in yellow and cyan). {\bf A.} Comparison for $N=4$. {\bf B.} Comparison for $N=100$; here, we used only the approximations $\ds \hat{\lambda}_{1,2}=N \pm \alpha N$, since the formal means are computationally too expensive.}}
\label{newman}
\end{figure}

\begin{figure}[h!]
\begin{center}
\includegraphics[width=.8 \textwidth]{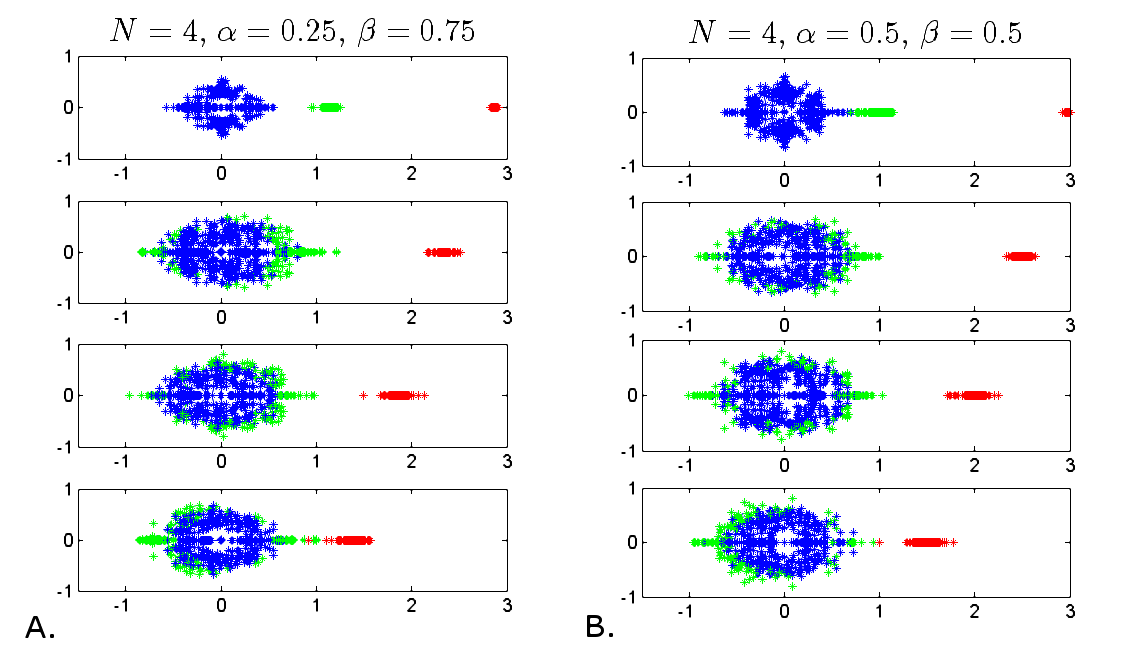}
\end{center}
\caption{\emph{ \footnotesize {\bf Illustration of separation of eigenvalues, when changing the community structure.} The eigenvalues are plotted in the complex plane: the leading eigenvalue in absolute value is shown in red, the second largest in green, the rest in blue. All plots are for $N=4$, and are based on samples of $100$ matrix configurations, under the following restrictions: {\bf A.} $\alpha=1/4$, $\beta=3/4$, $\gamma=1,3/4,1/2,1/4$ (from top to bottom). {\bf B.} $\alpha=1/2$, $\beta=1/2$, $\gamma=1,3/4,1/2,1/4$ (from top to bottom).}}
\label{wigner}
\end{figure}

\noindent so that $z_1>z_2$ if $\ds \alpha<1 - \frac{1}{N}$. In Figure~\ref{newman}, we show a comparison between our results and those of Nadakuditi and Newman, when applied to two fully-connected communities, by illustrating on the same axes $z_{1,2}$, the formal means $E({\lambda}_{1,2})$, and their close approximations obtained earlier as $\ds \hat{\lambda}_{1,2} = N \pm \alpha  N$. The approximations approach exactness in the large $N$ limit, at least for values of $\ds \alpha <1-\sqrt{\frac{2}{N}}$ (this is the density where $z_1$ has its global minimum, after which it shoots up, detaching from the graph of $E(\lvert {\lambda}_2 \rvert)$).

Lastly, the reference investigated the spectral distribution of the modularity matrix (i.e., whose spectral radius is, in the large $N$ limit, also the spectral radius for the adjacency matrix of the original non-oriented graph). The spectrum consisted of a continuous semicircular band of eigenvalues, and an additional, unique leading eigenvalue. As long as the leading eigenvalue is well separated from the semicircular band, there is evidence of community structure in the network; when the leading eigenvalue passes the edge of the band ($\ds z=\sqrt{c_\text{in}+c_\text{out}}$), the community structure is no longer detectable. As we have already suggested in Section ~\ref{subgraphs}, the property appears to extend to the case of the oriented random graph that constitutes our study case. In Figure~\ref{wigner} we show, for $N=4$, a sample ($100$ configurations) of the spectrum, observing the separation between eigenvalues, as the modularity of the network changes. Since the adjacency matrix is no longer symmetric, the eigenvalues are plotted in the complex plane. All eigenvalues are distributed within the unit disc, except the first two largest in absolute value, which, for $\gamma=1$, are real and significantly larger than 1. When beginning to decrease the ``community structure,'' (i.e., $\gamma$ decreases), the second leading eigenvalue collides into the unit disc, and starts diffusing around its boundary. If we continue decreasing $\gamma$, the first leading eigenvalue will also become indistinguishable from the pool distribution.



\subsection{Significance and applications to brain circuits}
\label{application}

Our basic results establish a connection between spectral properties of a bimodular oriented graph and the density of the inter-modular connections. While there are clearly better measures of architecture complexity in a network than edge density, our work was directly inspired by existing hypotheses that relate network functional efficiency precisely to the density of projections between subsets of network nodes. Our analysis is an attempt to provide a formal context and theoretical motivation for a multitude of existing empirical investigations, with the potential to reconcile results which may otherwise seem counter-intuitive, even self-contradictory.

For example, a body of evidence in the imaging and clinical literature relates emotional dysregulations (such as anxiety, depression, schizophrenia) to abnormal connectivity between the brain regions that regulate emotional responses. However, the results in the field seem ambiguous: some studies found that a lack of adequate amygdalar projections to prefrontal regions may be responsible for trait anxiety~\cite{kim2009structural,kim2011structural}, while other studies correlated the same phenomenon with major depression~\cite{dannlowski2009reduced}. A formal model investigating the effects of density on dynamics seemed therefore required to address these ambiguities, phrase the questions in the appropriate framework and reconcile the contradictions.

In our previous work~\cite{radulescu2013network}, we have used precisely the same graph-theoretical model as the one presented in this paper, in conjunction with nonlinear node dynamics, as a formal framework to study how network density can affect the complexity of signal outputs in a real brain network. Empirical time series were obtained using functional MRI in 96 human subjects with various types of emotional responses and anxiety levels. The brain system under study was the prefrontal-limbic meso-circuit (a feedback loop with well established contributions to emotion regulation), represented in our model as a network of excitatory and inhibitory nodes organized as two interconnected modules: the ``amygdala" (the excitatory component, responsible for emotional arousal), and the ``prefrontal cortex'' (the inhibitory unit, responsible for fear extinction). With each of the $N$ nodes in either module acting as a stochastic nonlinear oscillator (reflecting mean field behavior of neural populations), we studied how the level of connectivity between the two modules can determine and modulate efficiency of the system's dynamic responses. 

The optimality of responses was estimated from the scale-free features of the output signals. The scale-free behavior (which the model predicted accurately) was studied in both the empirical time series and in the model by calculating the fast Fourier transform for the discretized activations in each node, and then computing the slope of the best linear fit to the power spectrum in log-log coordinates. This measure, known as power spectral scale invariance (PSSI) is considered a straightforward way of characterizing complexity of a signal whose spectrum shows power-law behavior: $S(f) \sim f^{\beta_\text{PSSI}}$, by evaluating its relative frequency content. In this context, the scaling exponent $\beta_\text{PSSI}$ is 0 at maximum entropy (chaotic signals, also known as white-noise), and $\beta_\text{PSSI}$ close to  $-1$ and $-2$ represents increasing regularity and structure in the signals (known as pink and brown noise, respectively). To date, several studies have applied complexity analyses to fMRI, and have shown that for healthy neurobiological states, the entropy of neural time-series is characterized by roughly $\beta_\text{PSSI}= -1$, while neural time series in mental illnesses of systemic dysregulation (such as schizophrenia, anxiety, autism), show a significant shift towards $\beta_\text{PSSI} = 0$.

\begin{figure}[h!]
\begin{center}
\includegraphics[width = \textwidth]{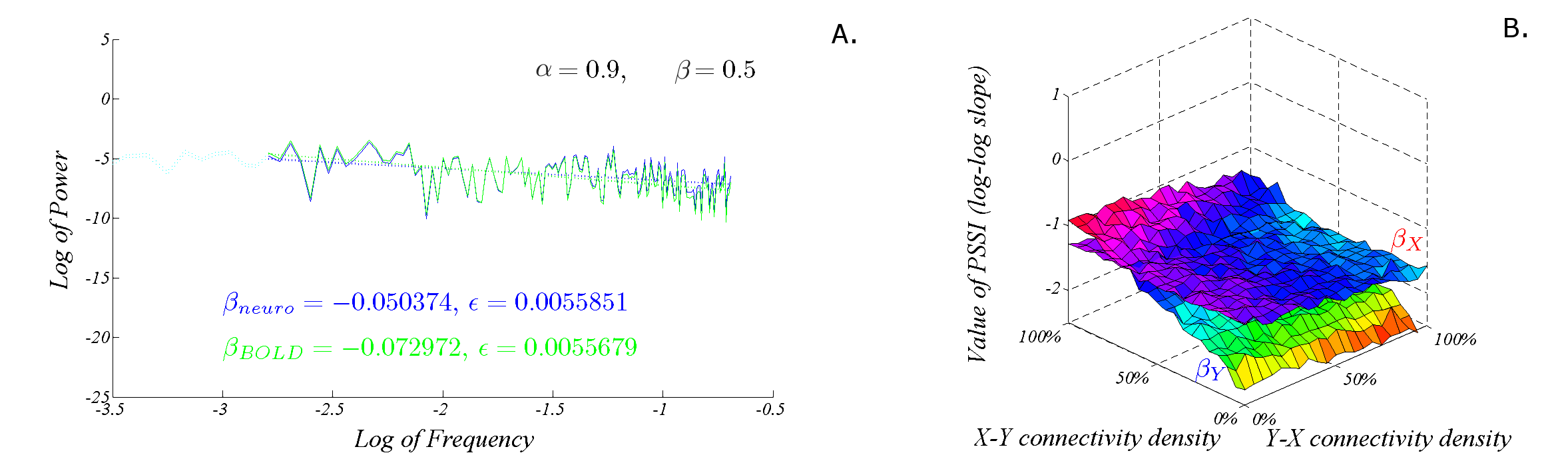}
\end{center}
\caption{ \label{figPSSI} \footnotesize \emph{ {\bf PSSI shifts as a function of input control and input density.} {\bf A.} Power spectra and best linear fit (with slopes shown in the legend), for the simulated neural power spectra (blue) and after applying the neuro-vascular model (green). {\bf B.} Dependence of $\beta_X$ (representing the average PSSI slope in module X, the amygdala) and $\beta_Y$ (representing the average PSSI slope in module $Y$, the prefrontal cortex), shown as surface functions of the $X$-to-$Y$ and $Y$-to-$X$ connectivity densities $\alpha$ and $\beta$. The simulations were performed for $N=20$ nodes in each module. The surfaces represent sample average slopes over all adjacency configurations with the given densities. Figure from the original manuscript}}

\vspace{3mm}
~\cite{radulescu2013network}.
\end{figure}

In our model, as in this paper, we allowed the excitatory and inhibitory connectivity densities $\alpha,\beta$ to vary within the interval $[0,1]$, and observed how the frequency profiles of the solutions (measured by the PSSI slopes) shifted from white to pink to brown noise. The results held qualitatively at the hemodynamic scale (modeled by introducing a neurovascular component), allowing us  to draw conclusions on how prefrontal-limbic connectivity may drive arousal dynamics and emotional responses, and helping us emit a testable hypothesis (see Figure~\ref{figPSSI}). Individuals with average emotional reactivity represent well-regulated control systems, in which excitatory (amygdala) and inhibitory (prefrontal) influences are balanced (these individuals exhibited fMRI signals close to pink noise in both amygdala and prefrontal regions). Anxious individuals have relatively weaker inhibitory feedback inputs from the prefrontal cortex (primarily driving amygdala signals closer to white noise). Less reactive individuals have relatively stronger excitatory inputs from the amygdala, producing stronger feedback (inducing more white noise primarily for the prefrontal cortex).

While this type of results are generally promising and clinically informative, one important step (and the center stone of our current work)  is to better understand their source. For example: an important, and rather surprising, feature of the model was that the local dynamics (as reflected by the PSSI values) were extremely robust between numerical runs (i.e., for different network configurations). Our two papers (the current one studying edge density based graph properties, the other -- studying their relationship with network dynamics~\cite{Chaos2015}), clarify that this is not a parameter-dependent property, or a numerical artifact -- but rather an intrinsic feature of the underlying graph. It is the robustness of certain network architectural features (in this case, the narrow distribution of the adjacency spectrum) that reflects into the robustness of the temporal systemic dynamics (as captured by the power spectra of the node trajectories).

\subsection{Adjacency to dynamics. Strengthening versus restructuring}
\label{adj_dynamics}

In ~\cite{Chaos2015}, we focus on how dynamic behavior depends on graph theoretical properties in nonlinear networks (i.e., the effects of changing the configuration of the network upon the temporal behavior of the system). As dynamics of coupled nonlinear oscillators have been widely investigated, it is has become clear that even trivial connectivity schemes, in conjunction with nonlinear behavior, may produce highly complex phenomena. 

For example, one of the most studied historical models in theoretical neuroscience (which has inspired many other analytical and modeling efforts~\cite{av1993basic,borisyuk1995dynamics}) is the Wilson-Cowan model~\cite{wilson1972excitatory} (a variation of the two-dimensional Fitzhugh-Nagumo equations), in which the coupled variables represent the fraction of neurons active at the current time in two excitatory and respectively inhibitory interacting populations. The model was shown to exhibit hysteresis and Hopf bifurcations, with bistability windows (in which the system has both an attracting equilibrium and an attracting limit cycle, separated by an unstable cycle). It has been later shown~\cite{borisyuk1995dynamics} that just by varying the strength of the symmetric weak coupling between two Wilson-Cowan excitatory/inhibitory units, one can produce very rich 4-dimensional phase-space transitions (bifurcations between symmetric, anti-symmetric and non-symmetric attractors like equilibria, cycles and invariant tori). 

We have studied the consequences of network architecture on Wilson-Cowan coupled dynamics~\cite{Chaos2015}, focusing primarily on finding the measures of architecture and dynamics that are optimal for quantifying their relationship. One interesting direction is to compare how dynamic behavior depends on architecture (viewed as a system parameter) versus how it depends on other parameters (see reference, as well as Appendix A). For example, consider two alternative ways to increase information diffusion between the two modules of our case study network: one by increasing the inter-modular edge \emph{ weights} and the other by increasing their \emph{ density}. Both actions lead to ``increasing connectivity'' between $X$ and $Y$, and to similar effects on the spectrum of the connectivity matrix, one may suspect that they also lead to similar changes in the temporal behavior of the corresponding dynamic network. However, our work suggests that this is not the case, and that the effects obtained when perturbing these two different aspects of the network connectivity can be very similar in some instances, but qualitatively different in others. 

Choosing the appropriate interplay between perturbing the configuration of the network and changing the coupling strengths seems to be an important part of the continuous choices a complex system like the brain needs to make to maintain optimal function. Under some circumstances, \emph{ local} configuration perturbations to the network may have more substantial dynamic effects than those obtained by a \emph{ global} change in the system's weights. In the context of optimal dynamics in a functional network, this may be seen as a vulnerability (simple addition of a few edges may drastically affect the function), but also as an adaptability feature (the system can more easily obtain the optimal flexibility which triggers efficient responses to the outside world).

\subsection{Edge updating and learning algorithms}

The oriented graph in this paper may be viewed as a representation for a network of coupled neurons, so that each edge represents a synapse with a corresponding ``weight,'' or synaptic strength (so that the connectivity matrix of such a network would represent the synaptic weight matrix). Synaptic updating has been well established as the physiological basis of learning, but the exact ways in which such a process is implemented biophysically are still under discussion.

There are many different models describing, qualitatively or quantitatively, the synaptic adjustments that may take place in a network of neurons during learning. In general, the process is assumed to involve not only weight changes of existing synapses, but also activation of ``silent'' sites (thus creating of new connections), and silencing, or pruning of active sites (thus deleting existing synapses). In terms of our model, this means that not only the edge weights, but also the edge distribution is likely to change during learning.

A clear biological restriction on synaptic updating has to be that the connections are somehow prevented from increasing without bound, which is why most models incorporate a normalization scheme. However, the manner in which a normalization step may actually be implemented by the brain is not at all clear, and has been subject of scientific controversy. Some rules assume the process to be local (e.g., subtractive normalization rules ~\cite{oja1982simplified,goodhill1993topography,willshaw1976patterned,miller1994role}, weight-dependent rules~\cite{elliott2002multiplicative} or BCM rules~\cite{cooper2004theory}), but one can imagine various other ways of insuring stability, possibly involving ``homeostasis'' or ``synaptic scaling''~\cite{turrigiano1998activity,turrigiano2004homeostatic}. Many models support a global normalization, for which the state of the whole network is assessed at each updating step, and a specific norm is imposed at each weight update.

In this light, it becomes important to understand the different consequences of using different normalization mechanisms when modeling synaptic updating and rewiring. While most models of learning introduce the updates into the weights themselves, the brain may additionally ``normalize'' (at least in the short term) by simply maintaining the overall \emph{ number} of active network connections approximately constant, so that, in the updating process, in the long term average one synapse will turn off whenever a new site is activated. One would then want to understand how these architectural dynamics may promote/influence learning, and how the effects of geometry updating complement or compare with the effects of direct weight updating. Let us finally note that the local mechanism of adding or deleting edges based on a probabilistic process (as described in Section ~\ref{general}), even though equivalent to our model in the large $N$ limit, produces substantially different spectra than our alternative normalization scheme for finite $N$ (see Figure~\ref{newman}b for $N=100$). Since many brain networks appear to operate with hundreds of nodes, it is important to understand the apparent distinctions between the two models, for relatively large, but finite values of $N$.

Knowledge of the geometry of the network is very important when determining which connectivity schemes are plausible to use for models of learning. The choices currently used in modeling range from considering fully-connected to fully-disconnected interacting modules, or layers~\cite{o2006making}. Our results suggest that convergence (learning) is not \emph{ a priori} prevented in either case. In developing future iterations of this model, it will also be important to explore how the learning process itself shapes the connectivity scheme. Siri et al.~\cite{siri2007effects} suggest that the structure emerging during learning breaks down into different numbers of hub-like subnetworks; this is very likely to affect the spectral robustness demonstrated in our modular network. Understanding the source and limits of this robustness is an instrument that could be used to investigate which architectures favor convergence under particular learning algorithms, and which not.

\bibliographystyle{plain}


\clearpage

\section*{Appendix A: Correspondence between adjacency spectra}, \\ \underline{Laplacian spectra and network dynamics}

In ~\cite{Chaos2015}, we considered the following 2N-dimensional system of coupled nonlinear oscillators:

\begin{eqnarray}
\dot{x}_k &=& \ds -x_k + (1-x_k) \cdot {\cal S}_{b_x,\theta_x} \left(-\sum_{p=1}^{N}{g_{yx} a_{kp} y_p} + \sum_{p=1}^{N}{g_{xx} x_p} + P \right) \nonumber \\
\dot{y}_k &=& \ds -y_k + (1-y_k) \cdot {\cal S}_{b_y,\theta_y} \left( \sum_{p=1}^{N}{g_{xy} b_{kp} x_p} + \sum_{p=1}^{N}{g_{yy} y_p} + Q \right)
\label{mothersys}
\end{eqnarray}

\noindent with $1 \leq k \leq N$. Each node is driven by external sources ($P$ for the nodes $x_k$ in the module $X$, and $Q$ for the nodes $y_k$ in the module $Y$). In addition, each node receives input from all other nodes that are connected to it through incoming edges, with weights $g$. The coefficients $a_{kp},b_{kp} \in \{ 0,1 \}$ are the binary entries of the adjacency blocks $A$ and $B$. The effective input to each node is the sum of all such external and internal sources, modulated by the sigmoidal:
\begin{equation}
{\cal S}_{b,\theta}[Z] = \frac{1}{1+\exp(-b[Z-\theta])}-\frac{1}{1+\exp(b\theta)}
\end{equation}

\noindent with parameters in the range used in the original Wilson-Cowan model~\cite{wilson1972excitatory}, as well as in subsequent papers~\cite{borisyuk1995dynamics}.

\begin{table}[h!]
\label{table_3_3}
\begin{center}
\begin{footnotesize}
\begin{tabular}{|c|c|c|c|}
\hline
& & &\\
$\;\left[ \begin{array}{cc|cc} &  & 1 & 1\\  &  & 1 & 0 \\ \cline{1-4} 1 & 1 &  & \\  1 & 0 &  & \end{array} \right] \; ({\cal A},I)_{iii}$ &
$\;\left[ \begin{array}{cc|cc} &  & 1 & 1\\  &  & 0 & 1 \\ \cline{1-4} 1 & 1 &  & \\  1 & 0 &  & \end{array} \right] \; ({\cal B,II})_{iv}$&
$\;\left[ \begin{array}{cc|cc} &  & 1 & 0\\  &  & 1 & 1 \\ \cline{1-4} 1 & 1 &  & \\  1 & 0 &  & \end{array} \right] \; ({\cal B,II})_{ii}$ &
$\;\left[ \begin{array}{cc|cc} &  & 0 & 1\\  &  & 1 & 1 \\ \cline{1-4} 1 & 1 &  & \\  1 & 0 &  & \end{array} \right] \; ({\cal C,III})_{i}$\\
& & &\\
\hline
& & &\\
$\;\left[ \begin{array}{cc|cc} & & 1 & 1\\  &  & 1 & 0 \\ \cline{1-4} 1 & 1 &  & \\  0 & 1 &  & \end{array} \right] \; ({\cal B},II)_{ii}$ &
$\;\left[ \begin{array}{cc|cc} & & 1 & 1\\  &  & 0 & 1 \\ \cline{1-4} 1 & 1 &  & \\  0 & 1 &  & \end{array} \right] \; ({\cal C},III)_{i}$ &
$\;\left[ \begin{array}{cc|cc} & & 1 & 0\\  &  & 1 & 1 \\ \cline{1-4} 1 & 1 &  & \\  0 & 1 &  & \end{array} \right] \; ({\cal A},I)_{iii}$ &
$\;\left[ \begin{array}{cc|cc} & & 0 & 1\\  &  & 1 & 1 \\ \cline{1-4} 1 & 1 &  & \\  0 & 1 &  & \end{array} \right] \; ({\cal B},II)_{iv}$ \\
& & &\\
\hline
& & &\\
$\;\left[ \begin{array}{cc|cc} & & 1 & 1\\  &  & 1 & 0 \\ \cline{1-4} 1 & 0 &  & \\  1 & 1 &  & \end{array} \right] \; ({\cal B},II)_{iv}$ &
$\;\left[ \begin{array}{cc|cc} & & 1 & 1\\  &  & 0 & 1 \\ \cline{1-4} 1 & 0 &  & \\  1 & 1 &  & \end{array} \right] \; ({\cal A},I)_{iii}$ &
$\;\left[ \begin{array}{cc|cc} & & 1 & 0\\  &  & 1 & 1 \\ \cline{1-4} 1 & 0 &  & \\  1 & 1 &  & \end{array} \right] \; ({\cal C},III)_{i}$ &
$\;\left[ \begin{array}{cc|cc} & & 0 & 1\\  &  & 1 & 1 \\ \cline{1-4} 1 & 0 &  & \\  1 & 1 &  & \end{array} \right] \; ({\cal B},II)_{ii}$ \\
& & &\\
\hline
& & &\\
$\;\left[ \begin{array}{cc|cc} & & 1 & 1\\  &  & 1 & 0 \\ \cline{1-4} 0 & 1 &  & \\  1 & 1 &  & \end{array} \right] \; ({\cal C},III)_{i}$ &
$\;\left[ \begin{array}{cc|cc} & & 1 & 1\\  &  & 0 & 1 \\ \cline{1-4} 0 & 1 &  & \\  1 & 1 &  & \end{array} \right] \; ({\cal B},II)_{ii}$ &
$\;\left[ \begin{array}{cc|cc} & & 1 & 0\\  &  & 1 & 1 \\ \cline{1-4} 0 & 1 &  & \\  1 & 1 &  & \end{array} \right] \; ({\cal B},II)_{iv}$ &
$\;\left[ \begin{array}{cc|cc} & & 0 & 1\\  &  & 1 & 1 \\ \cline{1-4} 0 & 1 &  & \\  1 & 1 &  & \end{array} \right] \; ({\cal A},I)_{iii}$\\
& & &\\
\hline
\end{tabular}
\end{footnotesize}
\end{center}
\caption{\label{table_3_3} \emph{ \footnotesize{\bf Classes of adjacency and Laplacian spectra in correspondence with dynamic classes, for N=2, density type $(\alpha,\beta)=(3/4,3/4)$.} Adjacency classes are designated as ${\cal A} - {\cal C}$, Laplacian classes as $I - III$, and dynamics classes by subscripts $i - iv$.}}
\end{table}

\noindent In the cited reference, we considered as an application networks of size 4 (i.e., $N=2$), and inspected the dynamic behavior of the system for every possible theoretical configuration of the adjacency matrix corresponding to a fixed pair of edge densities $(\alpha,\beta)$. To quantify the changes in dynamics produced by varying system parameters (such as, for example, the inter-modular connectivity weights $g_{xy}$ and $g_{yx}$), we used bifurcation diagrams in the $(g_{xy},g_{yx})$ parameter plane. Then, we observed how these diagrams changed when perturbing the underlying adjacency graph. We constructed all possible $(g_{xy},g_{yx})$ parameter planes that can be obtained for $N=2$ for each of two density pairs: $(\alpha,\beta)=(3/4, 3/4)$ and $(\alpha,\beta)=(1/2,3/4)$, respectively.  All 16 combinatorial configurations in ${\cal D}^{3/4,3/4}$  produced only four distinct dynamic parameter planes (which we labeled $i$ through $iv$). Similarly, all 24 combinatorial configurations in ${\cal D}^{1/2,3/4}$ produced only six dynamic classes (which we labeled $i$ through $vi$).

\begin{table}[h!]
\label{table_2_3}
\begin{footnotesize}
\begin{center}
\begin{tabular}{|c|c|c|c|}
\hline
& & & \\
$\;\left[ \begin{array}{cc|cc} &  & 1 & 1\\  &  & 0 & 0 \\ \cline{1-4} 1 & 1 &  & \\  1 & 0 &  & \end{array} \right] \; ({\cal A},I)_{v}$ &
$\;\left[ \begin{array}{cc|cc} & & 1 & 1\\  &  & 0 & 0 \\ \cline{1-4} 1 & 1 &  & \\  0 & 1 &  & \end{array} \right] \; ({\cal B},III)_{vi}$ &
$\;\left[ \begin{array}{cc|cc} & & 1 & 1\\  &  & 0 & 0 \\ \cline{1-4} 1 & 0 &  & \\  1 & 1 &  & \end{array} \right] \; ({\cal A},I)_{v}$ &
$\;\left[ \begin{array}{cc|cc} & & 1 & 1\\  &  & 0 & 0 \\ \cline{1-4} 0 & 1 &  & \\  1 & 1 &  & \end{array} \right] \; ({\cal B},III)_{vi}$\\
& & & \\
\hline
& & & \\
$\;\left[ \begin{array}{cc|cc} &  & 1 & 0\\  &  & 1 & 0 \\ \cline{1-4} 1 & 1 &  & \\  1 & 0 &  & \end{array} \right] \; ({\cal A},II)_{i}$ &
$\;\left[ \begin{array}{cc|cc} & & 1 & 0\\  &  & 1 & 0 \\ \cline{1-4} 1 & 1 &  & \\  0 & 1 &  & \end{array} \right] \; ({\cal A},II)_{i}$ &
$\;\left[ \begin{array}{cc|cc} & & 1 & 0\\  &  & 1 & 0 \\ \cline{1-4} 1 & 0 &  & \\  1 & 1 &  & \end{array} \right] \; ({\cal B},I)_{ii}$ &
$\;\left[ \begin{array}{cc|cc} & & 1 & 0\\  &  & 1 & 0 \\ \cline{1-4} 0 & 1 &  & \\  1 & 1 &  & \end{array} \right] \; ({\cal B},I)_{ii}$\\
& & & \\
\hline
& & & \\
$\;\left[ \begin{array}{cc|cc} &  & 1 & 0\\  &  & 0 & 1 \\ \cline{1-4} 1 & 1 &  & \\  1 & 0 &  & \end{array} \right] \; ({\cal C},III)_{iv}$ &
$\;\left[ \begin{array}{cc|cc} & & 1 & 0\\  &  & 0 & 1 \\ \cline{1-4} 1 & 1 &  & \\  0 & 1 &  & \end{array} \right] \; ({\cal D},I)_{iii}$ &
$\;\left[ \begin{array}{cc|cc} & & 1 & 0\\  &  & 0 & 1 \\ \cline{1-4} 1 & 0 &  & \\  1 & 1 &  & \end{array} \right] \; ({\cal D},I)_{iii}$ &
$\;\left[ \begin{array}{cc|cc} & & 1 & 0\\  &  & 0 & 1 \\ \cline{1-4} 0 & 1 &  & \\  1 & 1 &  & \end{array} \right] \; ({\cal C},III)_{iv}$\\
& & & \\
\hline
& & & \\
$\;\left[ \begin{array}{cc|cc} &  & 0 & 1\\  &  & 1 & 0 \\ \cline{1-4} 1 & 1 &  & \\  1 & 0 &  & \end{array} \right] \; ({\cal D},I)_{iii}$ &
$\;\left[ \begin{array}{cc|cc} & & 0 & 1\\  &  & 1 & 0 \\ \cline{1-4} 1 & 1 &  & \\  0 & 1 &  & \end{array} \right] \; ({\cal C},III)_{iv}$ &
$\;\left[ \begin{array}{cc|cc} & & 0 & 1\\  &  & 1 & 0 \\ \cline{1-4} 1 & 0 &  & \\  1 & 1 &  & \end{array} \right] \; ({\cal C},III)_{iv}$ &
$\;\left[ \begin{array}{cc|cc} & & 0 & 1\\  &  & 1 & 0 \\ \cline{1-4} 0 & 1 &  & \\  1 & 1 &  & \end{array} \right] \; ({\cal D},I)_{iii}$\\
& & & \\
\hline
& & & \\
$\;\left[ \begin{array}{cc|cc} &  & 0 & 1\\  &  & 0 & 1 \\ \cline{1-4} 1 & 1 &  & \\  1 & 0 &  & \end{array} \right] \; ({\cal B},I)_{ii}$ &
$\;\left[ \begin{array}{cc|cc} & & 0 & 1\\  &  & 0 & 1 \\ \cline{1-4} 1 & 1 &  & \\  0 & 1 &  & \end{array} \right] \; ({\cal B},I)_{ii}$ &
$\;\left[ \begin{array}{cc|cc} & & 0 & 1\\  &  & 0 & 1 \\ \cline{1-4} 1 & 0 &  & \\  1 & 1 &  & \end{array} \right] \; ({\cal A},II)_{i}$ &
$\;\left[ \begin{array}{cc|cc} & & 0 & 1\\  &  & 0 & 1 \\ \cline{1-4} 0 & 1 &  & \\  1 & 1 &  & \end{array} \right] \; ({\cal A},II)_{i}$\\
& & & \\
\hline
& & & \\
$\;\left[ \begin{array}{cc|cc} &  & 0 & 0\\  &  & 1 & 1 \\ \cline{1-4} 1 & 1 &  & \\  1 & 0 &  & \end{array} \right] \; ({\cal B},III)_{vi}$ &
$\;\left[ \begin{array}{cc|cc} & & 0 & 0\\  &  & 1 & 1 \\ \cline{1-4} 1 & 1 &  & \\  0 & 1 &  & \end{array} \right] \; ({\cal A},I)_{v}$ &
$\;\left[ \begin{array}{cc|cc} & & 0 & 0\\  &  & 1 & 1 \\ \cline{1-4} 1 & 0 &  & \\  1 & 1 &  & \end{array} \right] \; ({\cal B},III)_{vi}$ &
$\;\left[ \begin{array}{cc|cc} & & 0 & 0\\  &  & 1 & 1 \\ \cline{1-4} 0 & 1 &  & \\  1 & 1 &  & \end{array} \right] \; ({\cal A},I)_{v}$ \\
& & & \\
\hline
\end{tabular}
\end{center}
\end{footnotesize}

\caption{\label{table_2_3} \emph{ \footnotesize{\bf Spectral and dynamics classes for N=2, density type ($\alpha,\beta)=(1/2,3/4)$.} Adjacency classes are denoted ${\cal A} -- {\cal D}$, Laplacian classes are denoted $I -- III$, and dynamics classes are denoted as indeces as $i -- vi$.}}
\end{table}

In Tables~\ref{table_3_3} and ~\ref{table_2_3} we illustrate, for these two examples, to what extent cospectral and/or Laplacian cospectral graphs lead to the same dynamics. In the case of $(\alpha,\beta) = (3/4,3/4)$, there are three classes of  adjacency eigenspectra (designated by letters ${\cal A}$ through ${\cal C}$), which in this case are also the three classes for Laplacian eigenspectra (designated $I$ through $III$). The four distinct dynamics classes (designated by indices $i$ through $iv$) are mapped to the spectral classes in a well-defined, but not surjective way: that is, no dynamics can be obtained from multiple adjacency classes, but some adjacency classes can lead to multiple dynamics. 

Similarly, Table~\ref{table_2_3} shows how the six dynamic classes accessible to ${\cal D}^{1/2,3/4}$ are mapped to the adjacency and Laplacian spectral classes. In this case, the adjacency spectral classes (${\cal A}$ through ${\cal D}$) do not coincide with the Laplacian classes ($I$ through $III$). Dynamics is once again well-mapped to both adjacency and Laplacian spectral classes, although not surgectively (in fact, the many-to-one convergence is higher for Laplacian classes).

This suggests that, while the adjacency and Laplacian spectra, together with the density type, clearly have a contribution to dynamics, neither cannot be directly used to predict these dynamics. In fact, in this case, it is likely that the spectrum of the Laplacian gives less information on the dynamics than the spectum of the adjacency matrix.


\section*{Appendix B: Connecting sparser modules}
\label{subgraphs}

To investigate more general networks, we want to relax the full-connectedness condition of the two modules, and explore other intra-modular edge configurations, more realistic in the context of brain connectivity. As mentioned before, it is well known that the eigenspectrum of the adjacency matrix of a network organized in communities has leading eigenvalues that are well separated from the rest of the eigenvalues~\cite{chauhan2009spectral}. If our bimodular graph is thought of as describing the underlying coupling scheme for a dynamical system,  the position and overlap of the distributions ${\cal L}_j^{\alpha,\beta}$ will automatically reflect in the spectral properties of the network connectivity matrix (see Section~\ref{adj_dynamics}), and implicitly in the system's Jacobian matrix, thus affecting local dynamics around its equilibria.

 In this section, we illustrate in our specific case how the eigenvalue  distributions and the distance between them evolve as the modularity structure is gradually lost (how the leading eigenvalues approach the distribution of the remaining eigenvalues, as $\gamma$ decreases).

\begin{figure}[h!]
\begin{center}
\includegraphics[width=.9\textwidth]{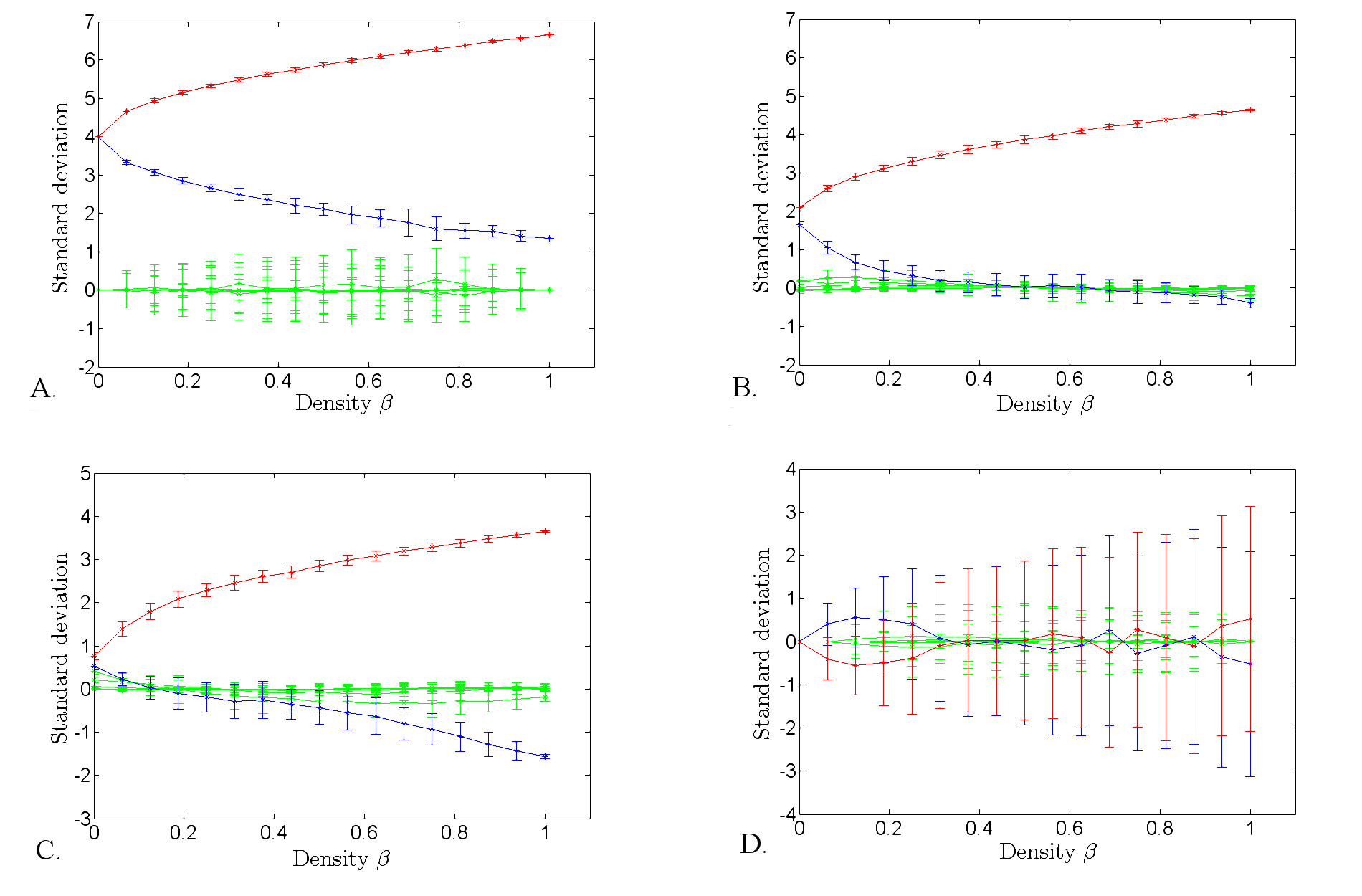}
\end{center}
\caption{ \emph{ \footnotesize {\bf Behavior of eigenvalues of ${\bf T}$} as $\gamma$ decreases from $1$ to $0$. Here, $N=4$, and $\alpha=1/2$. In each panel: $\gamma=1$ (panel {\bf A}), $\gamma=1/2$, (panel {\bf B}), $\gamma=1/4$ (panel {\bf C}) and $\gamma=0$ (panel {\bf D}). The mean values of each eigenvalue magnitude are represented along each curve together with the corresponding standard deviation (as error bars): the largest eigenvalue in red, the second largest in blue, and the remaining (footnotesize) eigenvalues in green.}}
\label{spread}
\end{figure}

Figure \ref{spread} shows the means and standard deviations of ${\cal L}_j^{\alpha,\beta}$ for four levels of intra-modular connectivity (each panel corresponds to a different value of $\gamma$, with $\alpha$ fixed and $\beta$ varied along the $x$-axis. When $\gamma=1$ (Figure \ref{spread}a), we recover the fully-connected modules studied in Section~\ref{adjacency}: the values of the standard deviations are small, and the first and second eigenvalues (whose means are well approximated in magnitude by $N \pm N \sqrt{\alpha \beta}$) remain to a large extent separated from the other small eigenvalues. As $\gamma$ decreases from $1$, this situation gradually changes, and the large expected eigenvalues decay in mean as $N \gamma \pm N\sqrt{\alpha \beta}$ respectively (Figure \ref{spread}b,c), to eventually completely collapse only when $\gamma=0$ (Figure \ref{spread}d).\\

To understand why the leading eigenvalues are close to  $N \gamma \pm N\sqrt{\alpha \beta}$ in mean, one can use a similar perturbation computation to the one carried out in Proposition~\ref{perturb}, as follows:


\begin{lemma} The matrix ${\bf C}=\left[ \begin{array}{c|c} \gamma {\bf M} & \alpha {\bf M} \\ \cline{1-2} \beta {\bf M} & \gamma {\bf M}  \end{array} \right]$ has eigenvalues:
\label{lemma2}
\begin{itemize}
\item $\lambda_1=\gamma N + N\sqrt{\alpha \beta}$, with corresponding eigenvector ${\bf u}_1= \left[ \begin{array}{c} \sqrt{a} \text{\large{\bf 1}} \\ \sqrt{b} \text{\large{\bf 1}} \end{array} \right]$
\item $\lambda_2=\gamma N - N\sqrt{\alpha \beta}$, with corresponding eigenvector ${\bf u}_2= \left[ \begin{array}{c} \sqrt{a} \text{\large{\bf 1}} \\ -\sqrt{b} \text{\large{\bf 1}} \end{array} \right]$
\item $\lambda_3=\hdots = \lambda_{2N}=0$, with corresponding eigenspace spanned by the vectors ${\bf u}_k=\left[ \begin{array}{c} {\bf t}_k \\ {\bf s}_k \end{array} \right]$, where ${\bf t}_k$ and ${\bf s}_k$ are $N \times 1$ column vectors with $\varphi({\bf t}_k)=\varphi({\bf s}_k)=0$, for $k \geq 3$.
\end{itemize}
\end{lemma}

\proof{ The proof follows directly from Lemma~\ref{lemma1}.}

\begin{prop}
The leading (real) eigenvalue of a binary matrix ${\bf T} = \left[ \begin{array}{c|c} {\bf P} & {\bf A} \\ \cline{1-2} {\bf B} & {\bf Q}  \end{array} \right]$, with $\varphi({\bf A})=\alpha N^2$, $\varphi({\bf B})=\beta N^2$ and $\varphi({\bf R})=\varphi({\bf S})=\gamma N^2$ is of the form:

$$\gamma N+N\sqrt{\alpha \beta} + \frac{1}{2N} \;\frac{1}{N+N\sqrt{\alpha \beta}} \; \text{\large{\bf 1}}^T {\bf W} \text{\large{\bf 1}} + O(1)$$

\noindent with 

$${\bf W} = {\bf R}^2 + {\bf S}^2+{\bf UV}+{\bf VU} +\frac{\alpha}{\beta} ({\bf VR}+{\bf SV}) + \frac{\beta}{\alpha} ({\bf RU}+{\bf US})$$ 

\noindent where ${\bf U}={\bf A}-\alpha{\bf M}$, ${\bf V}={\bf B}-\beta{\bf M}$, ${\bf R}={\bf P}-\gamma{\bf M}$ and ${\bf S}={\bf Q}-\gamma{\bf M}$ are all matrices with $\varphi({\bf U})=\varphi({\bf V})=\varphi({\bf R})=\varphi({\bf S})=0$.
\label{perturbation_sparse}
\end{prop}

\proof{The proof follows the same steps as Proposition~\ref{perturb}. As before, we consider a perturbation ${\bf T}_{\varepsilon} = {\bf C} + \varepsilon {\bf Z}$ of ${\bf C}$, where ${\bf Z} = \left[ \begin{array}{c|c} {\bf R} & {\bf U} \\ \cline{1-2} {\bf V} & {\bf S}  \end{array} \right]$ has $\varphi({\bf U})=\varphi({\bf V})=\varphi({\bf R})=\varphi({\bf S})=0$. We then similarly compute correction terms in the expansion of $\lambda_1$:

\begin{equation}
{\bf T}_\varepsilon ({\bf u}_1 + \varepsilon {\bf x}) = (\lambda_1+\varepsilon \mu_1)({\bf u}_1 + \varepsilon {\bf x})
\label{perturb2}
\end{equation}

\noindent with ${\bf x}$ perpendicular to ${\bf u}_1$. Expanding with respect to $\varepsilon$ and identifying the coefficients of $\varepsilon$, we get:

$${\bf Z u}_1 + {\bf Cx} = \lambda_1{\bf x} + \mu_1 {\bf u}_1$$

\noindent Expanding ${\bf x}=\sum x_k {\bf u}_k$, and ${\bf Z u}_j = \sum z_{jk} {\bf u}_k$ in the ${\bf C}$ eigenvector basis $({\bf u}_k)_{k=\overline{1,2N}}$ of eigenvectors of ${\bf C}$ and rewriting Equation~\eqref{perturb2} in components, we obtain that $\mu_1 = z_{11}$, and $\ds x_k = \frac{z_{1k}}{\lambda_1-\lambda_k}$, for $k \geq 2$.

\noindent We then calculate:

$${\bf Zu}_1 = \left[ \begin{array}{c|c} {\bf R} & {\bf U} \\ \cline{1-2} {\bf V} & {\bf S}  \end{array} \right] \left[ \begin{array}{c} \sqrt{a} \text{\large{\bf 1}} \\ \sqrt{b} \text{\large{\bf 1}} \end{array} \right] = \left[ \begin{array}{c} (\sqrt{a} {\bf R} + \sqrt{b} {\bf U})\text{\large{\bf 1}} \\ (\sqrt{a} {\bf V} + \sqrt{b} {\bf S})\text{\large{\bf 1}} \end{array} \right]$$

\noindent and, in components:

$${\bf Zu}_1 = z_{11} \left[ \begin{array}{c} \sqrt{a} \text{\large{\bf 1}} \\ \sqrt{b} \text{\large{\bf 1}} \end{array} \right] + z_{12} \left[ \begin{array}{c} \sqrt{a} \text{\large{\bf 1}} \\ -\sqrt{b} \text{\large{\bf 1}} \end{array} \right] + \sum_{k \geq 3} z_{1k} \left[ \begin{array}{c} {\bf t}_k \\ {\bf s}_k \end{array} \right]$$

\noindent Summing separately over the top and bottom $N$ entries, we get, respectively:

\begin{eqnarray}
N\sqrt{a}(z_{11} + z_{12}) &=& \sum_{k \geq 3} z_{1k}\varphi({\bf t}_k) = 0 \nonumber \\
N\sqrt{b}(z_{11} - z_{12}) &=& \sum_{k \geq 3} z_{1k}\varphi({\bf s}_k) = 0 \nonumber
\end{eqnarray}

\noindent implying that $z_{11}=z_{12}=0$, and subsequently $\mu_1=0$.\\

\noindent We continue for an $O(\varepsilon^2)$ approximation:

\begin{equation}
{\bf T}_\varepsilon ({\bf u}_1 + \varepsilon {\bf x} + \varepsilon^2 {\bf y}) = (\lambda_1+\varepsilon^2 \nu_1)({\bf u}_1 + \varepsilon {\bf x} + \varepsilon^2 {\bf y})
\label{perturb2}
\end{equation}

\noindent with ${\bf y}$ perpendicular to ${\bf u}_1$. Identifying the coefficients of $\varepsilon^3$ , we get:

\begin{eqnarray}
{\bf Cy} &+& {\bf Zx} = \lambda_1 {\bf y} + \nu_1{\bf u}_1 \nonumber
\end{eqnarray}

\noindent and, projected along the ${\bf u}_1$ component:

\begin{equation}
\nu_1 = \sum_k {\bf x}_k z_{k1} = \sum_{k \geq 2} \frac{z_{1k}z_{k1}}{\lambda_1-\lambda_k} \nonumber
\end{equation}

\noindent Since $z_{12}=0$ and $\lambda_k=0$, for $k \geq 3$, this becomes:

\begin{equation}
\nu_1 = \sum_{k \geq 3} \frac{z_{1k}z_{k1}}{\lambda_1} = \frac{1}{N+N\sqrt{\alpha \beta}} \sum_{k \geq 3} z_{1k}z_{k1} 
\end{equation}

\noindent If we expand ${\bf Z}^2$ in components as $\ds {\bf Z^2u}_1 = A_1{\bf u}_1+A_2{\bf u}_2 + \sum_{k \geq 3} A_k{\bf u_k}$, then $\sum_{k \geq 3} z_{1k}z_{k1} = A_1$. To calculate $A_1$, we calculate:

\begin{eqnarray}
\langle {\bf u}_1, {\bf Z}^2{\bf u}_1 \rangle = A_1 \| {\bf u}_1 \| + A_2 \langle {\bf u}_1, {\bf u}_2 \rangle = N(a+b)A_1 + N(a-b)A_2 \nonumber \\
\langle {\bf u}_1, {\bf Z}^2{\bf u}_1 \rangle = A_1 \langle {\bf u}_1, {\bf u}_2 \rangle + A_2 \| {\bf u}_2 \| = N(a-b)A_1 + N(a+b)A_2 \label{dot_product1}
\end{eqnarray}

\noindent Since ${\bf Z}^2 = \left[ \begin{array}{c|c} {\bf R}^2+{\bf UV} & {\bf RU}+{\bf US} \\ \cline{1-2} {\bf VS}+{\bf SV} & {\bf VU}+{\bf S}^2 \end{array} \right]$, we also have:

\begin{eqnarray}
\langle {\bf u}_1, {\bf Z}^2{\bf u}_1 \rangle = \text{\large{\bf 1}}^T [\alpha ({\bf R}^2+{\bf UV}) + \sqrt{\alpha \beta} ({\bf RU}+{\bf US}+{\bf VR}+{\bf SV}) + \beta ({\bf VU}+{\bf S}^2) ] \text{\large{\bf 1}} \nonumber \\
\langle {\bf u}_2, {\bf Z}^2{\bf u}_1 \rangle = \text{\large{\bf 1}}^T [\alpha ({\bf R}^2+{\bf UV}) + \sqrt{\alpha \beta} ({\bf RU}+{\bf US}-{\bf VR}-{\bf SV}) - \beta ({\bf VU}+{\bf S}^2) ] \text{\large{\bf 1}} \label{dot_product2}
\end{eqnarray}

\noindent Combining \eqref{dot_product1} and \eqref{dot_product2}, we get:

$\ds A_1 = \frac{1}{2N} \; \text{\large{\bf 1}}^T \left[{\bf UV}+{\bf VU}+{\bf R}^2+{\bf S}^2+\frac{\alpha}{\beta} \left({\bf VR}+{\bf SV} \right)+\frac{\beta}{\alpha} \left({\bf RU}+{\bf US} \right) \right] \text{\large{\bf 1}} $

\noindent hence

$\ds \nu_1 = \frac{1}{2N} \; \frac{1}{\gamma N+N\sqrt{\alpha \beta}} \; \text{\large{\bf 1}}^T \left[{\bf UV}+{\bf VU}+{\bf R}^2+{\bf S}^2+\frac{\alpha}{\beta} \left({\bf VR}+{\bf SV} \right)+\frac{\beta}{\alpha} \left({\bf RU}+{\bf US} \right) \right] \text{\large{\bf 1}}$

}


\vspace{1cm}
\noindent As before, however, this computation does not directly estimate the mean or the standard deviation of the eigenvalue distributions, which we will instead explore numerically below.

\begin{figure}[h!]
\begin{center}
\includegraphics[width=\textwidth]{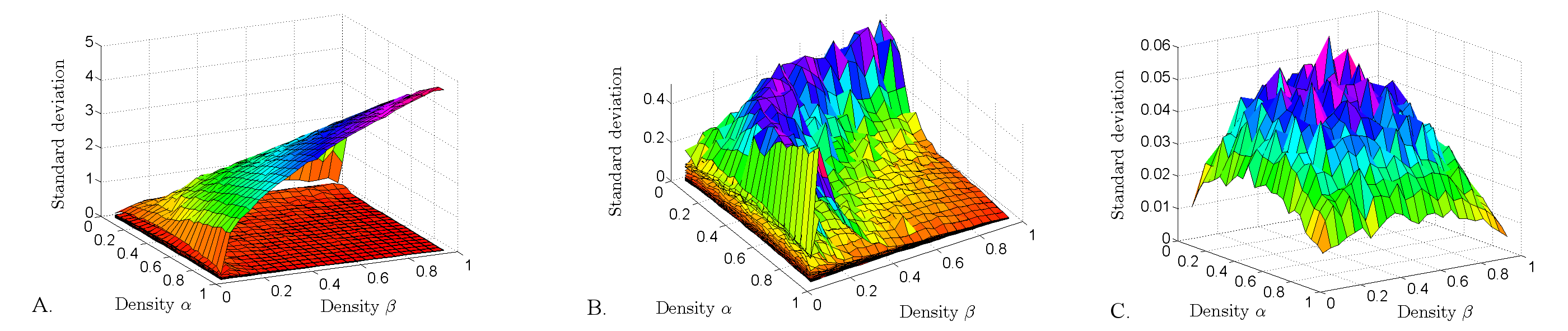}
\end{center}
\caption{ \emph{ \footnotesize {\bf Illustration, for $N=5$, of the standard deviation of ${\cal L}_1^{\alpha,\beta}$ for all values of intra-modular connectivity $\gamma$}. {\bf A.} Each surface represents the standard deviation of ${\cal L}_1^{\alpha,\beta}$, for one value of $\gamma \in [0,1]$ (low to high surfaces, as $\gamma$ decreases). For better visualization of the surfaces, we omitted the boundaries $(\alpha,1)$ and $(1,\beta)$. {\bf B.} The panel shows the same surfaces as in ({\bf A}), except that for the top surface corresponding to $\gamma=0$; this was excluded to better illustrate that, for all other values of $\gamma$, the standard deviations remain small, even with increasing $N$. {\bf C} The surface corresponding to $\gamma=1$ (shown in this panel) recovers the results in Section~\ref{adjacency} (compare with Figure~\ref{eigenvalues3}c, for $N=3$, with Figure~\ref{eigenvalues_8}b, for $N=8$, and with Figure~\ref{stdev}a, for multiple $N$ values). The computations were based on sample distributions obtained by considering for each $\gamma$ a sample of size $100$ pairs $({\bf P}, {\bf Q})$, and samples of size $10$ for ${\bf A}$ and for ${\bf B}$, for each fixed $\alpha$ and $\beta$.}}
\label{pruned}
\end{figure}

In Figure \ref{pruned}, we illustrate the dependence of the standard deviations simultaneously on the inter-modular edge densities $\alpha$ and $\beta$ (represented on the $x$ and $y$-axes), and on the intra-modular density $\gamma$ (different plots in each panel correspond to different values of $\gamma \in [0,1]$). The figure shows the standard deviation of ${\cal L}_1^{\alpha,\beta}$, for $N=5$, as a function of $(\alpha,\beta)$ for all discrete values $\gamma \leq 1$.  Let's notice first that, although the surfaces do not generally exhibit the same shape and unique ``central'' maximum as in the particular case of Section \ref{adjacency}, the unimodality still holds in cross-sections. Moreover, as $\gamma$ decreases, the standard deviation surfaces raise higher, corresponding to an expectable loss of the system's robustness when decreasing modular cohesion.  

However, while the standard deviation values do change with $\gamma$, the changes  do not appear to be all that significant until $\gamma$ actually approaches $0$. The values are instead bounded by a relatively small upper bound until $\gamma=0$, when this robustness breaks down. In the case of the leading eigenvalue, the depreciation is monotonous: the standard deviations, very small when $\gamma=1$, increase slowly as $\gamma$ decreases from $1$, then faster as the values of $\gamma$ get close to $0$, with a complete crash occurring at $\gamma=0$ (also see Figure \ref{spread}). For inter-modular connectivity close to saturation (i.e.,  pairs $(\alpha,\beta)$ close to the corner $(1,1)$), the surfaces are barely affected by the intra-modular density $\gamma$, as long as $\gamma > 0$. If one had speculated that the intra-modular full-connectedness confers robustness to the network eigenvalue spectrum, one would now notice that this robustness is surprisingly well preserved as the foll-connectedness is gradually loosened, by pruning out random edges and thus lowering the intra-modular density. The property is completely lost only when the two moduli remain totally disconnected. We further interpret this in the Discussion.

\begin{figure}[h!]
\begin{center}
\includegraphics[width=\textwidth]{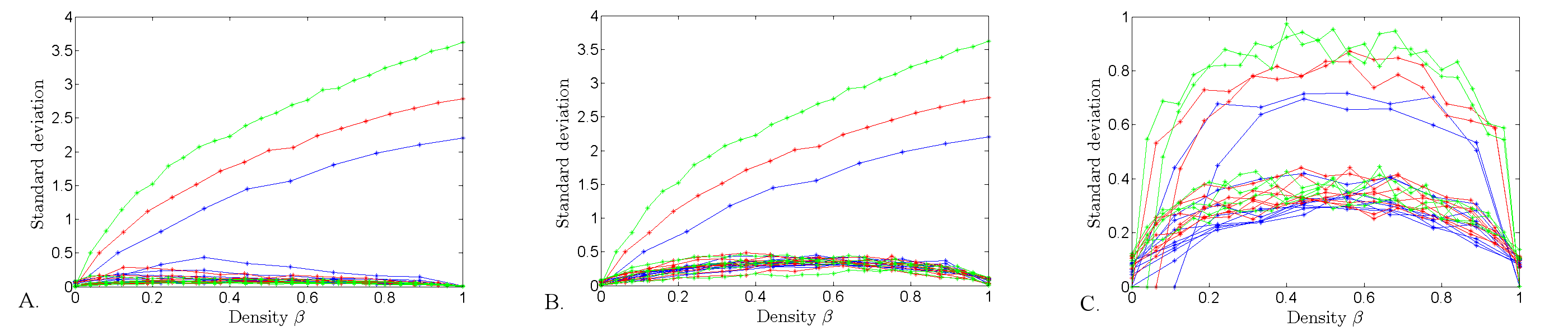}
\end{center}
\caption{ \emph{ \footnotesize {\bf Behavior of the standard deviations when increasing the size $N$}. Shown in blue are the curves for $N=3$, corresponding to all possible values of $gamma=k/9$ for $0 \leq k \leq 9$. In red are the curves for $N=4$, corresponding to $\gamma=2k/16$, for $0 \leq k \leq 8$. In green are the curves for $N=5$, corresponding to $\gamma=5k/25$, for $0 \leq k \leq 5$ . For ${\cal L}_1^{\alpha,\beta}$ and ${\cal L}_2^{\alpha,\beta}$ (panels {\bf A} and {\bf B}), the standard deviations are low, expect in the extreme case $\gamma=0$ (top curve of each color). For ${\cal L}_3^{\alpha,\beta}$, the standard deviations remain low for all $\gamma$, with a slight increase with $N$ for values of $\gamma$ close to zero (top two curves of each color). The computations were based on the same sample distributions as were used for Figure~\ref{pruned}.}}
\label{gamma_N}
\end{figure}

The next natural question is to ask, as before, how the robustness of the distributions changes with the size $N$. In Figures~\ref{gamma_N} b,c and d we show cross-sections of the surfaces introduced in Figure~\ref{pruned} (obtained by fixing one density $\alpha$), compared for increasing values of $N$, suggesting that robustness is not substantially affected when the network increases in size, except for values of $\gamma$ close to zero.

\end{document}